\documentstyle[aps,prd,amssymb]{revtex}
\input epsfig.sty
\def\laq{\raise 0.4ex\hbox{$<$}\kern -0.8em\lower 0.62 ex\hbox{$\sim$}}
\def\gaq{\raise 0.4ex\hbox{$>$}\kern -0.7em\lower 0.62 ex\hbox{$\sim$}}
\def\dis{\displaystyle}
\def\bdis{\begin{displaymath}}
\def\edis{\end{displaymath}}
\newcommand{\beq}{\begin{equation}}
\newcommand{\eeq}{\end{equation}}
\newcommand{\beqn}{\begin{eqnarray}}
\newcommand{\eeqn}{\end{eqnarray}}
\newcommand{\nbeqn}{\begin{eqnarray*}}
\newcommand{\neeqn}{\end{eqnarray*}}
\newcommand{\bcen}{\begin{center}}
\newcommand{\ecen}{\end{center}}
\newcommand{\bfig}{\begin{figure}}
\newcommand{\efig}{\end{figure}}
\newcommand{\btab}{\begin{table}}
\newcommand{\etab}{\end{table}}
\newcommand{\btabu}{\begin{tabular}}
\newcommand{\etabu}{\end{tabular}}
\newcommand{\benu}{\begin{enumerate}}
\newcommand{\eenu}{\end{enumerate}}
\newcommand{\bite}{\begin{itemize}}
\newcommand{\eite}{\end{itemize}}
\newcommand{\bary}{\begin{array}}
\newcommand{\eary}{\end{array}}

\newcommand{\ud}{\mathrm{d}}

\def\dis{\displaystyle}
\def\lsim{\raise 0.4ex\hbox{$<$}\kern -0.8em\lower 0.62
ex\hbox{$\sim$}}
\def\gsim{\raise 0.4ex\hbox{$>$}\kern -0.7em\lower 0.62
ex\hbox{$\sim$}}

\newcommand{\mpl}{M_{\rm Pl}}

\draft

\begin{document}

\title{Upgraded VIRGO detector(s) and stochastic gravitational waves 
backgrounds}

\author{ D. Babusci$^{(a)}$ 
\footnote{Electronic address: danilo.babusci@lnf.infn.it }  
and M. Giovannini$^{(b)}$\footnote{
Electronic address: Massimo.Giovannini@ipt.unil.ch} }

\address{{\it $^{(a)}$ INFN - Laboratori Nazionali di 
Frascati, 00044 Frascati, Italy}}

\address{{\it $^{(b)}$ Institute for Theoretical Physics, Lausanne University,
BSP-Dorigny, CH-1015 Switzerland }}

\maketitle
\begin{abstract}
The sensitivity achievable by a pair of VIRGO detectors to 
stochastic and isotropic gravitational wave 
backgrounds of cosmological origin is discussed 
in view of the development of a second VIRGO interferometer. 
We describe a semi-analytical  
technique allowing to compute the signal-to-noise ratio for 
(monotonic or non-monotonic) 
logarithmic energy spectra of relic gravitons of arbitrary
 slope. We apply our results to the case of two correlated and 
coaligned VIRGO detectors  and we compute their achievable 
sensitivities. The maximization of the overlap 
reduction function is discussed. 
We focus our attention on a class of models whose 
expected sensitivity is more promising, namely the case of string 
cosmological gravitons. We perform our calculations both for 
the case of minimal string cosmological scenario and in the case of 
a non-minimal scenario where a long dilaton dominated phase is 
present prior to the onset of the ordinary 
radiation dominated phase. In this framework, we study possible 
improvements of the achievable sensitivities by selective 
reduction of the thermal contributions 
(pendulum and pendulum's internal modes)
to the noise power spectra of the detectors. 
Since a reduction of the shot noise does not increase significantly 
the expected sensitivity of a VIRGO pair (in spite of the 
relative spatial location of the two detectors)
our findings support the experimental efforts directed towards a substantial 
reduction of thermal noise.

\end{abstract}

\renewcommand{\theequation}{1.\arabic{equation}}
\setcounter{equation}{0}

\section{The problem and its motivations}

It is well known that every variation of the background 
geometry produces graviton pairs which are stochastically 
distributed and whose logarithmic energy spectra represent 
a faithful snapshot of the (time) evolution of the curvature 
scale at very early times \cite{1}. Indeed, one of the 
peculiar features of stochastic graviton backgrounds is that 
their energy spectra extend over a huge interval of (present)
frequencies. This feature can be appreciated by comparing 
the graviton backgrounds with other backgrounds of 
electromagnetic origin (like the cosmic microwave background 
[CMB]). The analysis of the CMB background (together with its spatial 
anisotropies) is relevant for very large (length) scales 
\cite{2} (roughly ranging between the present horizon 
[i.e. $10^{-18} $ Hz] and the horizon at decoupling 
[i.e. $ 10^{- 16 } $ Hz]). Since gravitational interactions 
are much weaker than electromagnetic interactions they also 
decouple much earlier and, therefore, the logarithmic energy 
spectra of relic gravitons produced by the pumping action of 
the gravitational field can very well extend for (approximately) 
twenty five orders of magnitude in frequency \cite{3}. From the 
physical point of view, this observation implies that the energy 
spectra of relic gravitons can be extremely relevant in order to 
probe the past history of the Universe in a regime which will 
never be directly accessible with observations of electromagnetic 
backgrounds. 

In spite of the fact that the {\em nature} 
of the production mechanism is shared by different 
types of models \cite{1}, the specific {\em amplitudes of the 
energy spectra} can very well change depending 
upon the behavior of the background evolution. 
An example in this direction are 
logarithmic energy spectra increasing in frequency \cite{9}.
Different theoretical signals (with different 
spectral distributions) lead to 
 detector outputs of different amplitudes. 
We are facing a non-linear problem where a change in the detector signal
can be determined either by an improvement 
in the features of the detector or by a different 
functional form of the logarithmic energy spectrum \cite{noi}.
Therefore, in order to evaluate the performances 
of a given detector one has to choose the specific 
functional form of the logarithmic energy spectrum. 
A possible choice is represented 
by scale invariant spectra \cite{gri,8}. Another 
rather interesting choice is represented 
by tilted (``blue'' \cite{gri2}) spectra whose energetical content 
is typically concentrated at frequencies larger than 
the mHz \cite{gio2}. 
String cosmological models \cite{10} are yet another interesting 
theoretical laboratory leading usually to sizable theoretical 
signals in the operating window of wide band interferometers (WBI) 
\cite{11}. A possible 
detection of these backgrounds would represent an 
interesting test for cosmological models 
inspired by the low energy string effective action. 

Every measurement in cosmology 
turns out to be difficult for different and independent  
reasons. The CMB anisotropy experiments have to cope 
with  the mandatory subtraction of different 
electromagnetic foregrounds which can be much 
larger than the ``cosmological" signal one 
ought to detect. In order to detect gravitational 
waves of cosmological origin with terrestrial 
measurements we are facing similar problems.

The signal induced in the detector output by stochastic 
gravitational waves 
backgrounds is indistinguishable from the intrinsic noise of the 
detector itself. This implies that, unless the amplitude of the signal 
is very large, 
the only chance of direct detection of these backgrounds
lies in the analysis of the 
correlated fluctuactions of the outputs of, at least, two detectors 
affected by independent noises. The problem of the optimal processing 
of the detector outputs required for the detection of the stochastic 
background has been considered by various authors \cite{mic,chr} 
and it was also reviewed in ref. \cite{alr}. 

Suppose, indeed, that the  signal registered at each detector can be 
written as (we limit ourselves to the case of two detectors 
$(i\,=\,1,2)$)
\begin{equation}
s_{i}\,=\,h_{i}(t)\,+\,n_{i}(t)\,,
\end{equation} 
where we have indicated with $n$ the intrinsic noise of the detector, 
and with $h$ the gravitational strain due to the stochastic background. 
By assuming that the detector noises are stationary and uncorrelated, 
the ensemble average of their Fourier components satisfies
\begin{equation}
\langle n^{\ast}_{i}(f)\,n_{j}(f')\rangle\,=\,\frac{1}{2}\,\delta(f-f')
\,\delta_{ij}\,S^{(i)}_{n}(|f|)\;, 
\end{equation}
where $S_{n}(|f|)$ is usually known as the one-sided noise power spectrum 
and is expressed in seconds. Starting to the signals $s_1$ and $s_2$, a 
correlation ``signal'' for an observation time $T$ can be defined in the 
following way:
\begin{equation}
S\,=\,\int_{- T/2}^{T/2}\,{\rm d} t\,\int_{- T/2}^{T/2}\,{\rm d} t'
\,s_1 (t)\,s_2 (t')\,Q (t - t')
\end{equation} 
where $Q$ is a filter function that depends only by $t - t'$ because we 
assume that $n$ and $h$ are both stationary. The optimal choice of $Q$ 
corresponds to the maximization of the signal-to-noise ratio associated 
to the ``signal'' $S$. Under the further assumptions that detector noises 
are Gaussian, much larger in amplitude than the gravitational strain and 
statistically independent on the strain itself, it can be shown 
\cite{mic,chr,alr} that the signal-to-noise ratio in a frequency range 
$(f_{\rm m},f_{\rm M})$ is given by 
\footnote{In order to avoid possible confusions we stress that
 the definition of 
the SNR is the one discussed in \cite{noi} and it is essentially the square 
root of the one discussed in \cite{mic,chr,alr}.}:
\begin{equation}
{\rm SNR}^2 \,=\,\frac{3 H_0^2}{2 \sqrt{2}\,\pi^2}\,F\,\sqrt{T}\,
\left\{\,\int_{f_{\rm m}}^{f_{\rm M}}\,{\rm d} f\,
\frac{\gamma^2 (f)\,\Omega_{{\rm GW}}^2 (f)}{f^6\,S_n^{\,(1)} (f)\,
S_n^{\,(2)} (f)}\,\right\}^{1/2}\; ,
\label{SNR}
\end{equation}
where $H_0$ is the present value of the Hubble parameter and $F$ is a 
numerical factor depending upon the geometry of the two detectors. 
In the case of the correlation between two interferometers $F= 2/5$, 
however, in the correlation of detectors of different geometry, 
$F\neq 2/5$ (see Appendix A for details about this point).
In Eq. (\ref{SNR}), 
the  performances achievable by the pair of detectors are certainly  
controlled by the noise power spectra (NPS) $S_n^{\,(1,2)}$. 
However in Eq. (\ref{SNR}), 
on top of NPS, there are two important quantities. 
The first one is the {\em theoretical} 
background signal defined through the logarithmic energy spectrum 
(normalized to the critical density $\rho_c$) and expressed at the present 
(conformal) time\footnote{In most of our equations we drop the 
dependence of spectral quantities upon the present time since all the 
quantities introduced in this paper are evaluated today.} $\eta_0$
\begin{equation}
\Omega_{{\rm GW}}(f,\eta_0)\,=\,\frac{1}{\rho_{c}}\,
\frac{{\rm d} \rho_{{\rm GW}}}{{\rm d} \ln{f}}\,=\, 
\overline{\Omega}(\eta_0)\,\omega(f,\eta_0)\,.
\label{1}
\end{equation}
The second one is the overlap reduction function $\gamma(f)$ \cite{chr,alr}
which is a dimensionless function describing the reduction in  the sensitivity
of the two detectors (at a given frequency $f$)
arising from the fact that the two detectors are not in the 
same place and, in general, not coaligned
(for the same location and orientation $\gamma (f) = 1$). Since 
the overlap reduction function cuts-off 
the integrand of Eq. (\ref{SNR}) at a frequency which approximately 
corresponds to the inverse separation between the two detectors, it may  
represent a dangerous (but controllable) element in the reduction 
of the sensitivity of a given pair of detectors.

Various ground-based interferometric detectors are presently under 
construction (GEO \cite{geo}, LIGO-LA, LIGO-WA \cite{lig}, TAMA \cite{tam}, 
VIRGO \cite{vir}). Among them,
 the pair consisting of most homogeneous (from the point of view 
of the noise performances) detectors with minimum separation is given by the 
two LIGOs (VIRGO and GEO are even closer, but they have different performances 
for what concerns the NPS). However, this separation 
($\simeq$ 3000 km) is still too large. The overlap 
reduction function $\gamma (f)$ for the pair 
LIGO-LA$-$LIGO-WA encounters 
its first zero at 64 Hz,  falling off (swiftly) at higher 
frequencies, i.e., right in the region where the two LIGOs, at least in 
their initial version, have better noise performances.

Recently, within the european gravitational wave community, 
the possibility of building in Europe an interferometric 
detector of dimensions comparable to VIRGO has  received 
close attention \cite{gia}. Therefore, there is a chance 
that in the near future the VIRGO detector, 
now under construction at Cascina (Pisa) in Italy, will be complemented 
by another interferometer of even better performances very close 
(at a distance $d\,<\,1000$ km) to it. In 
this paper we examine in detail the possible improvements in the VIRGO 
sensitivity as a result of direct correlation of two VIRGO-like detectors. 
Furthermore, since technological improvements in the construction of the 
interferometers can be reasonably expected in the next years, it is easy 
to predict that also VIRGO, as for the LIGO detectors, will gradually 
evolve toward an advanced configuration. For this reason we also examine 
the possible consequences of a selective improvements of the noise 
characteristics of the two detectors on the obtained results.

In order to evaluate precisely the performances of 
a pair of VIRGO detectors we will use the following 
logic. First of all we will pick up a given class
of theoretical models which look particularly 
promising in view of their spectral properties 
in the operating window of the WBI. 
Secondly we will analyze the signal-to-noise 
ratios for different regions of the parameter
space of the model. Finally we will 
implement some selective reduction of the noises 
and we will compare the results with the ones 
obtained in the cases 
where the noises are not reduced. We will 
repeat the same procedure for different classes 
of models. 

The results and the investigations we are 
reporting can be applied to spectra of arbitrary 
functional form. The only two requirements 
we assume will be the continuity of the 
logarithmic energy spectra (as a function of the present 
frequency) and of their first derivative. We will also give 
some other examples in this direction.

In order to make our analysis 
concrete we will pay particular attention to the evaluation 
of the performances of a pair of VIRGO detectors in the case 
of string cosmological models \cite{10,11}.

The plan of our paper is then the following. In Section II
we introduce the basic semi-analytical tecnique which allows 
the evaluation of the SNR for a pair of WBI. In Section III 
we will evaluate the performances of a pair of VIRGO detectors 
in the case of string cosmological models.
In Section IV we will show how to implement 
a selective noise reduction and we will investigate the 
impact of such a reduction in the case 
of the parameter space of the models previously analyzed. 
Section V contains our final discussion and the 
basic summary of our results. 
We collect in the Appendices some  technical results useful for our analysis 
and other interesting complements to our investigation.

\renewcommand{\theequation}{2.\arabic{equation}}
\setcounter{equation}{0}
\section{SNR evaluation}

In the operating window of the VIRGO detectors the theoretical 
signal will be defined through the logarithmic energy spectrum 
reported in Eq. (\ref{1}). In the present Section we shall not make any 
specific assumption concerning $\omega(f)$ and our results have 
general applicability. We will only assume that it is a continuous 
function of the frequency and we will also assume that its first 
derivative is well defined in the operating window of WBI. 
This means that $\omega(f)$ can be, in principle, a non-monotonic 
function.

\subsection{Basic Formalism}

\begin{figure}
\centerline{\epsfxsize = 7 cm  \epsffile{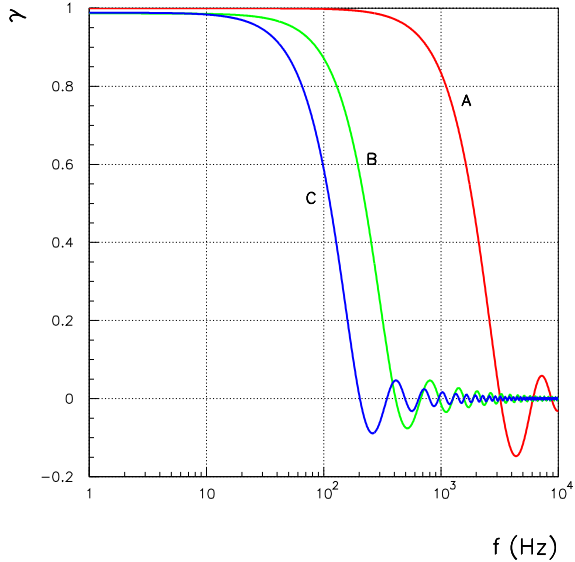}} 
\vspace*{-0.5cm}
\caption[a]{We report the overlap reduction function(s) for the correlation of 
the VIRGO detector presently under construction in Cascina 
(43.6 N, 10.5 E) with a coaligned interferometer whose (corner) 
station is located at: A) (43.2 N, 10.9 E), $d\,=\,58$ km (Italy); 
B) (43.6 N, 4.5 E), $d\,=\,482.7$ km (France); C) (52.3 N, 9.8 E), 
$d\,=\,958.2$ km (Germany). The third site (C) corresponds
to the present location of  the GEO detector. Notice that from A to C 
the position of the first zero of $\gamma(f)$ gets shifted in the infra-red. 
See also Appendix A concerning this last point.}
\label{over}
\end{figure}

The noise power spectrum of the VIRGO detector is well approximated by 
the analytical fit of Ref. \cite{cuo}, namely 
\begin{equation}
\Sigma_n (f)\,=\,\frac{S_n (f)}{S_0}\,=\,
\left\{
\begin{array}{lc}
\infty & \qquad \qquad f < f_b \\ [8pt]
\dis \Sigma_1\,\biggl(\frac{f_{{\rm a }}}{f}\biggr)^5\,+\,
\dis \Sigma_2\,\biggl(\frac{f_{{\rm a}}}{f}\biggr)\,+\,
\dis \Sigma_3\,\biggl[ 1 + \biggl(\frac{f}{f_{\rm a}}\biggr)^2\biggr],& 
\qquad \qquad f \ge f_b
\label{NPS}
\end{array}
\right.
\end{equation}
where 
\bdis
S_0\,=\,10^{-44}\,{\rm s}\;,\qquad f_a\,=\,500\,{\rm Hz}\;, 
\qquad f_b\,=\,2\,{\rm Hz}\;,\qquad
\begin{array}{l}
\Sigma_1\,=\,3.46\,\times\,10^{-6} \\
\Sigma_2\,=\,6.60\,\times\,10^{-2} \\ 
\Sigma_3\,=\,3.24\,\times\,10^{-2}\,.
\end{array}
\edis
In order to compute reliably (and beyond naive power counting 
arguments) the SNR we have to specify the overlap reduction 
function $\gamma(f)$. The relative location and orientation of the 
two detectors determines the functional form of $\gamma(f)$ 
which  has to be gauged in such a way that the 
overlap between the two detectors 
is maximized (i.e. $\gamma(f) \simeq 1$ for most of the 
operating window of the two VIRGO). Moreover, the two interferometers
of the pair 
should also be sufficiently far apart in order to decorrelate the 
local seismic and electromagnetic noises. Since the precise 
location of the second VIRGO detector has not been specified so 
far \cite{gia}, we find useful to elaborate about this point by computing 
the overlap reduction functions corresponding to two coaligned 
VIRGO interferometers with different spatial separations. The 
results of these calculations are reported in Fig. \ref{over}. 
Needless to say that these choices are purely theoretical and are 
only meant to illustrate the effects of the distance 
on the performances of the VIRGO pair\footnote{For illustrative 
purposes, we assumed that a distance of about 50 km is sufficient 
to decorrelate local seismic and e.m. noises. Such a  hypothesis is 
fair  at the present stage  and it is certainly justified within the 
spirit of our exercise. However, at the moment, we do not 
have any indication either against or in favor of our choice.}. 
The curves labeled with A, B, and C shown in Fig. \ref{over} 
correspond to different distances $d$ between the site 
of the VIRGO detector (presently under construction 
in Cascina, near Pisa) and the central corner station 
of a second coaligned VIRGO interferometer. Let us now look at 
the position of the first frequency $f_i$ for which   
$\gamma(f_i)\,=\,0$ for each of the curves. We can notice 
that by increasing  $d$ (i.e., going from A to C) 
the value of $f_i$ gets progressively shifted towards lower and lower 
frequencies,  linearly with $d$. This 
means that, for the specific purpose of the detection 
of a stochastic gravitational waves background, the position 
of the first zero of the overlap reduction function cannot 
be ignored. On a general ground we would like $f_i$ to be 
slightly larger than the frequency region where the sensitivity 
of the pair of wide band detectors is maximal.  In the explicit examples 
presented in this paper we will focus our attention 
on the case A. The other two 
configurations will be the subject of a related investigation \cite{bg}.

\subsection{SNR versus phenomenological bounds on the graviton spectrum}

By inserting the parametrization (\ref{1}) into  Eq. (\ref{SNR}) 
we can write 
\begin{equation}
{\rm SNR}^2 \,=\,\frac{3 H_0^2}{5 \sqrt{2}\,\pi^2}\;
\sqrt{T}\;\frac{\overline{\Omega}}{f_0^{5/2}\,S_0}\;J \;,
\label{snrrescaled}
\end{equation}
where we introduced the (dimension-less) integral 
\begin{equation}
J^2 \,=\,\int_{\nu_{\rm m}}^{\nu_{\rm M}}\,{\rm d} \nu\,
\frac{\gamma^2\,(f_0 \nu)\,\omega^2(f_0 \nu)}
{\nu^6\,\Sigma_n^{\,(1)} (f_0 \nu)\,
\Sigma_n^{\,(2)} (f_0 \nu)}\;.
\label{Jint}
\end{equation}
Here the integration variable is $\nu\,=\,f/f_0$, with $f_0$ a 
generic frequency scale within the operating window of the 
interferometer, and the integration domain is restricted to 
the region $f_{\rm m}\,\le\,f\,\le\,f_{\rm M}$ 
(i.e., $\nu_{\rm m}\,\le\,\nu\,\le\,\nu_{\rm M}$). 
In the following we will choose $f_0\,=\,100$ Hz and, taking into 
account the frequency behavior of $\gamma (f)$ (see Fig. \ref{over}), 
we can assume  $f_{\rm M}\,=\,10$ kHz (i.e., $\nu_{\rm M}\,=\,100$). 
The lower extreme $f_{\rm m}$ is put equal to the frequency $f_b$ 
entering Eq. (\ref{NPS}) (i.e., $\nu_{\rm m}\,=\,0.02$).

For the chosen values of $f_0$ and $S_0$ (see Eq. (\ref{NPS})) 
one has:
\begin{equation}
h_{0}^2\,\overline{\Omega}\,\simeq\,\frac{4.0\,\times\,10^{-7}}{J}\;
\left(\,\frac{1\;{\rm yr}}{T}\,\right)^{1/2}\;{\rm SNR}^2\;.
\label{sens}
\end{equation}
Since we will often refer to this formula we want to 
stress its physical meaning. Suppose that the functional form of $\omega (f)$
is given. Then the numerical value of the 
integral $J$ can be precisely computed and, through Eq. (\ref{sens}), 
$\overline{\Omega}$ can be estimated. 
This quantity, inserted in Eq. (\ref{1}), 
determines for each frequency $f$ the minimum $\Omega_{\rm GW}$ 
detectable (for an observation time $T$, with a signal-to-noise 
ratio SNR) by the correlation of the two detector outputs.

In the next section, $\overline{\Omega}$ will be compared with 
two other quantities: $\overline{\Omega}^{\,{\rm th}}$ and 
$\overline{\Omega}^{\,{\rm max}}$. The first is the theoretical 
value of the normalization of the spectrum, while the second 
represents the largest normalization compatible with the 
phenomenological bounds applicable to the stochastic GW 
backgrounds. These quantities are of different nature and
in order to be more precise let us consider an example.
 
Suppose, for simplicity, that we are dealing with a 
logarithmic energy spectrum which is a monotonic function 
of the present frequency. Suppose, moreover, that the spectrum 
decreases sufficiently fast in the infra-red in order 
to be compatible both with the pulsar timing bound and with the 
CMB anisotropies bounds. Then the most relevant bound will come,
effectively, from Big-Bang nucleosynthesis (BBN) \cite{sch,wal,cop}.
Therefore, in this particular case, we will have that 
$\overline{\Omega}^{\,{\rm max}}$ is determined by demanding 
that 
\begin{equation}
h^2_0\,\int\,\Omega_{\rm GW}(f,\eta_0)\;{\rm d}\ln{f}\,<\,0.2\,
h_0^2\,\Omega_{\gamma}(\eta_0)\,\simeq\,5\,\times\,10^{-6},
\label{ns}
\end{equation}
where $\Omega_{\gamma}(\eta_0)\,=\,2.6\,\times\,10^{-5}\,h_0^{-2}$
is the present fraction of critical energy density stored in radiation.
According to our definition, $\overline{\Omega}^{\,{\rm max}}$ is
the maximal normalization of the spectrum compatible with the previous 
inequality, namely,
\begin{equation}
h_0^2\,\overline{\Omega}^{\,{\rm max}}\,\simeq\,
\frac{5\,\times\,10^{-6}}{{\cal I}}\;, \qquad 
{\cal I}\,= \,\int_{f_{\rm ns}}^{f_{\rm max}}\,
\omega(f)\;{\rm d}\ln{f}.
\label{NSnorm}
\end{equation}
Notice that  $f_{\rm ns}\,\simeq\,10^{-10}$ Hz is the present value 
of the frequency corresponding to the horizon at the nucleosynthesis 
time; $f_{\rm max}$ stands for  the maximal frequency of the spectrum 
and it depends, in general, upon the specific theoretical model.
If the spectrum has different slopes, $\overline{\Omega}^{\,{\rm max}}$ 
will be determined not only by the nucleosynthesis bound but also by 
the combined action of the CMB anisotropy bound \cite{2,ben} and 
of the pulsar timing bound \cite{kas}. Indeed, we know that 
the very small fractional timing error in the arrival times of 
the millisecond pulsar's pulses implies that 
$\Omega_{{\rm GW}}\,\laq\,10^{-8}$ for a frequency which is roughly 
comparable with the inverse of the observation time along which 
pulsars have been monitored 
(i.e., $\omega_{\rm p}\,\sim\,1/T_{\rm obs}\,=\,10^{-8}$ Hz). Moreover, 
the observations of the large scale anisotropies in the microwave sky 
\cite{ben} imply that the graviton contribution to the integrated 
Sachs-Wolfe effect has to be smaller than (or at most of the 
order of) the detected amount of anisotropy. This observation implies 
that $\Omega_{{\rm GW}}\,\leq\,6.9\,\times\,10^{-11}$ for 
frequencies ranging between the typical frequency of the present 
horizon and a frequency thirty of forty times larger. In the case of 
a logarithmic energy density with decreasing slope the 
$\overline{\Omega}^{\,{\rm max}}$ will be mainly determined by the 
Sachs-Wolfe bound and it will be the maximal normalization of the 
spectrum compatible with such a bound. 

On a general ground, we will have that 
$\overline{\Omega}^{\,{\rm th}}\,\leq\,\overline{\Omega}^{\,{\rm max}}$, 
namely the theoretical normalization of the spectrum is bounded, from 
above, by the maximal normalization compatible with all the 
phenomenological bounds. Therefore, the mismatching 
 between these  quantities can be interpreted as an 
effective measure of the theoretical error 
in the determination of the absolute normalization of the spectrum.

Since $\omega(f)$ enters (in a highly non-linear way) into the form 
of $J$ (as defined in Eq. (\ref{Jint})), the corresponding 
$\overline{\Omega}$ in Eq. (\ref{sens}) will be different for any 
(specific) frequency dependence in $\omega(f)$. The consequence 
of this statement is that it is not possible to give a general 
(and simple) relationship between the sensitivity at a given frequency,
the spectral slope and the (generic) theoretical amplitude of the 
spectrum. However, given the form of the theoretical spectrum, the 
phenomenological bounds (depending upon the theoretical slope) 
will fix uniquely the theoretical error and the maximal achievable 
sensitivity. So, if we want to evaluate the performances of the VIRGO 
pair we should pick up a given class of theoretical models (characterized 
by a specific functional form of $\omega(f)$) and compute 
the corresponding sensitivity. The same procedure should then be repeated 
for other classes of models and, only at the end, the 
respective sensitivities can be compared.

\renewcommand{\theequation}{3.\arabic{equation}}
\setcounter{equation}{0}
\section{Primordial graviton background versus VIRGO*VIRGO}

We can consider, in principle, logarithmic energy spectra with 
hypothetical analytical forms and arbitrary normalizations. 
If the logarithmic energy spectrum is either a flat or a decreasing 
function of the present frequency \cite{gri}, we can expect, 
on general grounds, that the theoretical signal will be of the 
order of (but smaller than) $10^{-15}$ \cite{rub} for present 
frequencies comparable with the operating window of the 
VIRGO pair. This happens because of the combined action of the 
Sachs-Wolfe bound together with the spectral behavior of ther 
infra-red branch of the spectrum produced thanks to the 
matter-radiation transition. 
Of course this observation holds for  models where the graviton 
production occurs because of the adiabatic variation 
of the background geometry \footnote{ An exception to this 
 assessment is represented by cosmic strings models 
 leading to a flat logarithmic energy spectrum for frequencies 
 between $10^{-12}$ Hz and $10^{-8}$ Hz \cite{vac}. 
Another possible exception is 
 given by the gravitational power radiated by magnetic
 (and hypermagnetic) \cite{gio1}
 knot configurations at the electroweak scale \cite{sha1}.}.
 
In order to have large signals falling in the operating window of 
the VIRGO pair we should have deviations from scale invariance 
for frequencies larger than few mHz. Moreover, 
these deviations should go in the direction of increasing 
logarithmic energy spectra. This is what happens  
in the case of quintessential inflationary models \cite{gio2}. 
In this case, however, as we discussed in a previous 
analysis \cite{noi}, the BBN bound put strong constraints on the 
theoretical signal in the operating window of the 
VIRGO pair.

Another class of model leading to a large theoretical 
signal for frequencies between few Hz and 10 kHz  is 
represented by string cosmological models \cite{9,10,11}. 
Therefore, in order
to evaluate the performances of the VIRGO pair 
and in order to implement a procedure of selective noise 
reduction we will use string cosmological spectra.

\subsection{Minimal models of pre-big-bang}

In string cosmology 
and, more specifically, in the pre-big-bang scenario, the curvature scale
and the dilaton coupling are both growing in cosmic time. Therefore 
the graviton spectra will be {\em increasing} in frequency  instead of 
{\em decreasing} as it happens in ordinary inflationary models. 

In the context of string cosmological scenarios the Universe 
starts its evolution in a very weakly coupled regime with vanishing 
curvature and dilaton coupling. After a phase of sudden 
growth of the 
curvature and of the coupling the corrections to the tree level action 
become important and the Universe enters a true stringy phase which is 
followed by the ordinary radiation dominated phase. It should be stressed 
that the duration of the stringy phase is not precisely known 
and it could happen that all the physical scales contained 
within our present Hubble radius crossed the horizon during the stringy phase
as pointed out in \cite{gas1}.

The maximal amplified frequency of the graviton spectrum is given 
by \cite{9,11}
\begin{equation}
f_{1}(\eta_0)\,\simeq\,64.8\,\sqrt{g_1}\,\left(\,\frac{10^{3}}
{n_r}\,\right)^{1/12}\; {\rm GHz}
\end{equation}
where $n_{r}$ is the effective number of spin degrees 
of freedom in thermal equilibrium at the end of the stringy phase,
and $g_1\,=\,M_{s}/\mpl$ where $M_s$ and $\mpl$ are the string 
and Planck masses, respectively. Notice that $g_1$ is the value 
of the dilaton coupling at the end of the stringy phase, and 
is typically of the order of $10^{-2}\,\div\,10^{-1}$ \cite{kap}.
In order to red-shift the maximal amplified frequency of the 
spectrum from the time $\eta_1$ (which marks the beginning of 
the radiation dominated evolution) up to the present time we 
assumed that the cosmological evolution prior to $\eta_0$ and 
after $\eta_1$ is adiabatic. Minimal models of pre-big-bang are 
the ones where a dilaton dominated phase is followed by a stringy 
phase which terminates at the onset of the radiation dominated 
evolution. In the context of minimal models, the function 
$\omega (f)$ introduced in Eq. (\ref{1}) can be written as 
\begin{equation}
\omega (f)\,=\,
\left\{
\begin{array}{lc}
\dis z_s^{- 2 \beta}\,\left(\,\frac{f}{f_s}\,\right)^3\,\left[\,1\,+\,
z_s^{2 \beta - 3}\,-\,\frac12\,\ln{\frac{f}{f_s}}\,\right]^2 
& \dis \qquad \qquad f\,\le\,f_s\,=\,\frac{f_1}{z_s} \\ [15pt]
\dis \left[\,\biggl(\frac{f}{f_1}\biggr)^{3 - \beta}\,+\,
\biggl(\frac{f}{f_1}\biggr)^{\beta}\,\right]^2 
& \qquad \qquad f_s\,<\,f\,\le\,f_1
\end{array}
\right.
\label{minth}
\end{equation}
where, 
\begin{equation}
\dis \beta\,=\,\frac{\ln\,(g_1/g_s)}{\ln\,z_s}.
\end{equation}
In this formula $z_s\,=\,f_1/f_s$ and $g_s$ are, respectively, the 
red-shift during the string phase 
and the value of the coupling constant at the end of the dilaton dominated 
phase.
The first of the two branches appearing in Eq. (\ref{minth}) 
is originated by modes leaving the horizon 
during the dilaton dominated phase and re-entering 
during the radiation dominated phase. The second branch is 
mainly originated by modes leaving the horizon during the 
stringy phase and re-entering always in the radiation 
dominated phase.
The  theoretical normalization 
\begin{equation}
\overline{\Omega}^{\,\rm th}\,=\,2.6\,g_1^2\,\left(\,\frac{10^3}{n_r}\,
\right)^{1/3}\,\Omega_{\gamma}(\eta_0)\;,
\label{omth}
\end{equation}
multiplied by  $\omega(f)$ (as given in Eq. (\ref{minth})) leads 
to the theoretical form of the spectrum.
Notice that $n_r$ is of the order of $10^2\,\div\,10^{3}$ (depending 
upon the specific string model) and it represents a theoretical 
uncertainty.

However, as anticipated in the previous section, the theoretical 
normalization of the spectrum should be contrasted with the one 
saturating the BBN bound (i.e., $\overline{\Omega}^{\,{\rm max}}$). 
This quantity is obtained by Eq. (\ref{NSnorm}), where in the case 
under consideration
\begin{equation}
{\cal I}\,=\,{\cal I}_{d}\,+\,{\cal I}_{s} \qquad {\rm with}
\qquad {\cal I}_{d}\,=\,\int_{f_{\rm ns}}^{f_s}\,\frac{{\rm d}f}{f}\,
\omega (f)\;,\qquad {\cal I}_{s}\,=\,\int_{f_{s}}^{f_1}\,
\frac{{\rm d}f}{f}\,\omega (f)\;.
\end{equation}
The analytical expressions of ${\cal I}_{d}$ and ${\cal I}_{s}$ are 
reported in Appendix B. We have to bear in mind that in the intermediate 
frequency region of the graviton spectra an important bound comes 
from the pulsar timing measurements. Therefore, if one ought to 
consider rather long stringy phases (i.e., large $z_s$), the BBN 
constraint should be supplemented by the requirement that 
$\Omega_{\rm GW}(10^{-8}\,{\rm Hz})\,<\,10^{-8}$ \cite{kas}. We will come 
back to this point later.

Following the explicit expression of the function $\omega (f)$, 
Eq. (\ref{sens}) can be re-written as follows:
\begin{equation}
h_0^2\,\overline{\Omega}\,\simeq\,4\,\times\,10^{-7}\,\left(\,
\frac{1\,{\rm yr}}{T}\,\right)^{1/2}\,
\frac{{\rm SNR}^2}{\sqrt{J_d^2\,+\,J_s^2}},
\end{equation}
where, introduced the following notation
\begin{eqnarray}
J_{k} &=& \int_{\nu_{\rm m}}^{\nu_{s}}\,{\rm d}\nu\,
\frac{\gamma^2 (f_0 \nu)}{\Sigma_{n}^{(1)} (f_0\nu)\,
\Sigma_{n}^{(2)} (f_0\nu)}\,\ln^{k}{\nu}\;,
\qquad k\,=\,0,1,2,3,4 \nonumber\\[10pt]
J_{\pm m (3 - 2 \beta)} &=& \int_{\nu_s}^{\nu_{\rm M}}\,
{\rm d}\nu\,\frac{\gamma^2 (f_0 \nu)}{\Sigma_{n}^{(1)} (f_0 \nu)\,
\Sigma_{n}^{(2)} (f_0\nu)}\,\nu^{\pm m (3 - 2 \beta)}\;,
\qquad m\,=\,1,2 \\[10pt]
C_d &=& 1\,+\,z_s^{2 \beta - 3}\,+\,\frac12\,\ln{\nu_s}\;, \nonumber
\end{eqnarray}
one has
\begin{eqnarray}
J_{d} &=& \frac{z_s^{3 - 2 \beta}}{\nu_1^3}\,
\left(\,C_d^4 J_0\,-\,2 C_{d}^3 J_1\,+\,\frac32 C_{d}^2 J_2\,-\,
\frac12 C_{d} J_3\,+\,\frac{1}{16} J_{4}\,\right)^{1/2}\;,
\nonumber \\[10pt]
J_{s} &=& \frac{1}{\nu_1^3}\,\left(\,6 J_0\,+\,
\frac{J_{6 - 4 \beta}}{\nu_1^{6 - 4 \beta}}\,+\, 
\frac{J_{4 \beta - 6}}{\nu_1^{4 \beta - 6}}\,+\, 
4 \frac{J_{3 - 2 \beta}}{\nu_1^{3 - 2 \beta}}\,+\, 
4 \frac{J_{2 \beta - 3}}{\nu_1^{2 \beta - 3}}\,\right)^{1/2}\;.
\end{eqnarray}

The previous expressions are general in the sense that they are
applicable for a generic value of $f_s$. 
If $f_{\rm m}\,<\,f_{s}\,<\,f_{\rm M}$ then both $J_{s}$ and $J_{d}$ 
give contribution to the sensitivity. If, on the other hand 
$f_s\,<\,f_{\rm m}$ (i.e., a long stringy phase) the main contribution 
to the sensitivity will come from $J_{s}$. The integrals appearing in 
$J_{d,s}$ have to be evaluated numerically. In all our calculations we 
will assume that both VIRGO detectors are characterized by the same 
(rescaled) NPS (reported in Eq. (\ref{snrrescaled})). 

The main steps of our calculation are the following. We firstly fix $g_1$ and 
for each pair $(z_s,\,g_1/g_s)$ (within the range of their physical 
value) we compute $\overline{\Omega}$  (for $T\,=\,1$ yr and SNR = 1), 
and $\overline{\Omega}^{\,{\rm max}}$. We then compare these two quantities 
to the theoretical normalization given in Eq. (\ref{omth}). If 
$\overline{\Omega}^{\,{\rm th}}$ will be larger than $\overline{\Omega}$ 
(but smaller than $\overline{\Omega}^{\,{\rm max}}$) 
we will say that the theoretical 
signal will be ``visible'' by the VIRGO pair. In this way we will 
identify in the plane $(z_s,\,g_1/g_s)$ a visibility region according 
to the sensitivity of the VIRGO pair. The theoretical error on the border 
of this region can be estimated by substituting 
$\overline{\Omega}^{\,{\rm max}}$ to $\overline{\Omega}^{\,{\rm th}}$.

To illustrate this point we consider a specific case. The value of the 
coupling at the end of the stringy phase can be estimated to lie between 
0.3 and 0.03 \cite{kap}. 
The knowledge of $g_1$ will not fix uniquely the theoretical 
spectrum which does also depend on the number of relativistic degrees of 
freedom at the end of the stringy phase. Therefore, the theoretical error 
in the determination of the absolute normalization of the spectrum could 
be also viewed as the error affecting the determination of $n_r$. 
In all the plots shown we will take, when not otherwise stated, 
$g_1\,=\,1/20$ and $n_{r}\,=\,10^3$ as fiducial values. Different choices 
of $g_1$ will lead to similar results. We will also 
assume that the overlap reduction function associated with the 
pair is the one reported in the curve A of Fig. \ref{over}. 

In Fig. \ref{minrat} (top left) we report the result of our 
calculation for the ratio between 
$\overline{\Omega}^{\,{\rm max}}$ and $\overline{\Omega}$ as a 
function of $g_1/g_s$ and $\log{z_s}$. The contour plot (bottom left) 
shows the region of the plane $(\log{z_s},\,g_1/g_s)$ 
where this ratio is greater than 1, i.e. the maximal visibility 
region allowed by the BBN bound. In the opposite case, i.e., 
$\overline{\Omega}^{\,{\rm max}}/\overline{\Omega}\,<\,1$, 
the VIRGO pair is sensitive to a region excluded by the BBN.
In the right part of Fig. \ref{minrat} we go one step further and 
we plot the ratio between $\overline{\Omega}^{\,{\rm th}}$ and 
$\overline{\Omega}$. The shaded area in the 
contour plot (bottom right) is the region of the plane $(\log{z_s},\,g_1/g_s)$ 
where the conditions 
$\overline{\Omega}^{\,{\rm th}}/\overline{\Omega}\,>\,1$ and 
$\overline{\Omega}^{\,{\rm max}}/\overline{\Omega}\,>\,1$ are 
simultaneously met. The shaded area in this plot defines the visibility 
region of the VIRGO pair {\em assuming} the theoretical normalization 
of the spectrum. From  Fig. \ref{minrat}, 
by ideally subtracting the shaded area of the left contour 
plot from the shaded area of the right contour plot
we obtain an estimate of the theoretical error.
The results we just presented can be obviously recovered for different
values of $g_1$ close to one. However, if $g_1$ gets too small 
(and typically below 1/25) the visibility area gets smaller and 
smaller eventually disappearing. 

The visibility regions appearing in Fig. \ref{minrat} extend from 
intermediate values of $z_s$ (of the order of $10^{8}$) towards 
large values of $z_s$ (of the order of $10^{18}$). Notice that 
for our choice of $g_1$, $f_s$ can become as small as $10^{-8}$
for $z_s $ of the order of $10^{18}$. As we recalled in the 
previous Section, this frequency corresponds to the inverse of 
the observation time along which pulsar signals have been monitored 
and, therefore, for this  frequency, a further ``local'' bound 
applies to the logarithmic energy spectra of relic gravitons. 
This bound implies that 
$\Omega_{\rm GW}(10^{-8}\,{\rm Hz})\,<\,10^{-8}$. In our examples, 
the compatibility with the BBN bound implies also that the 
pulsar timing constraint is satisfied. Given our choice 
for $g_1$ we can clearly see that the visibility regions depicted 
in Fig. \ref{minrat} extend for values of $g_s$ which can be as 
small as 1/160 (or as small as 1/60 in the case of right part of 
Fig. \ref{minrat}).

\subsection{Non-minimal models of pre-big-bang}

In the context of minimal models of pre-big-bang, the 
end of the stringy phase coincides with the 
onset of the radiation dominated evolution.
At the moment of the transition to the 
radiation dominated phase the dilaton seats 
at its constant value. This 
means that $g_1\,\sim\,0.03\,\div\,0.3$ at the beginning 
of the radiation dominated evolution.
As pointed out in \cite{gas1}, it is  not be impossible to imagine 
a scenario where the coupling constant is still 
growing while the curvature scale starts decreasing in time.
In this type of scenario the stringy phase is followed 
by a phase where the dilaton still increases, or, in other 
words, the coupling constant is rather small at the moment 
where the curvature starts decreasing so that $g_1\,\ll\,1$. 
\begin{figure}
\begin{center}
\begin{tabular}{cc}
      \hbox{\epsfxsize = 8 cm  \epsffile{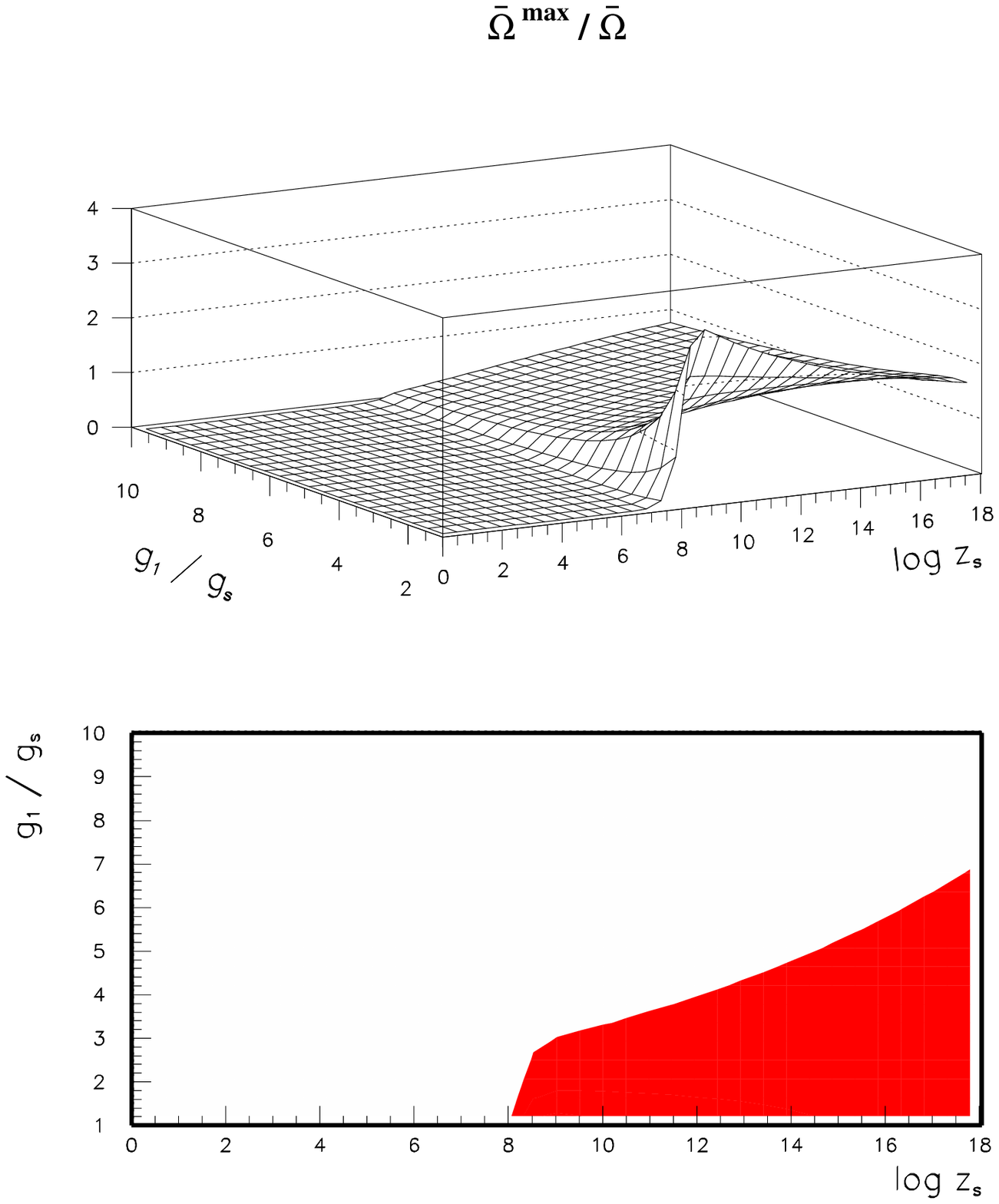}} &
      \hbox{\epsfxsize = 8 cm  \epsffile{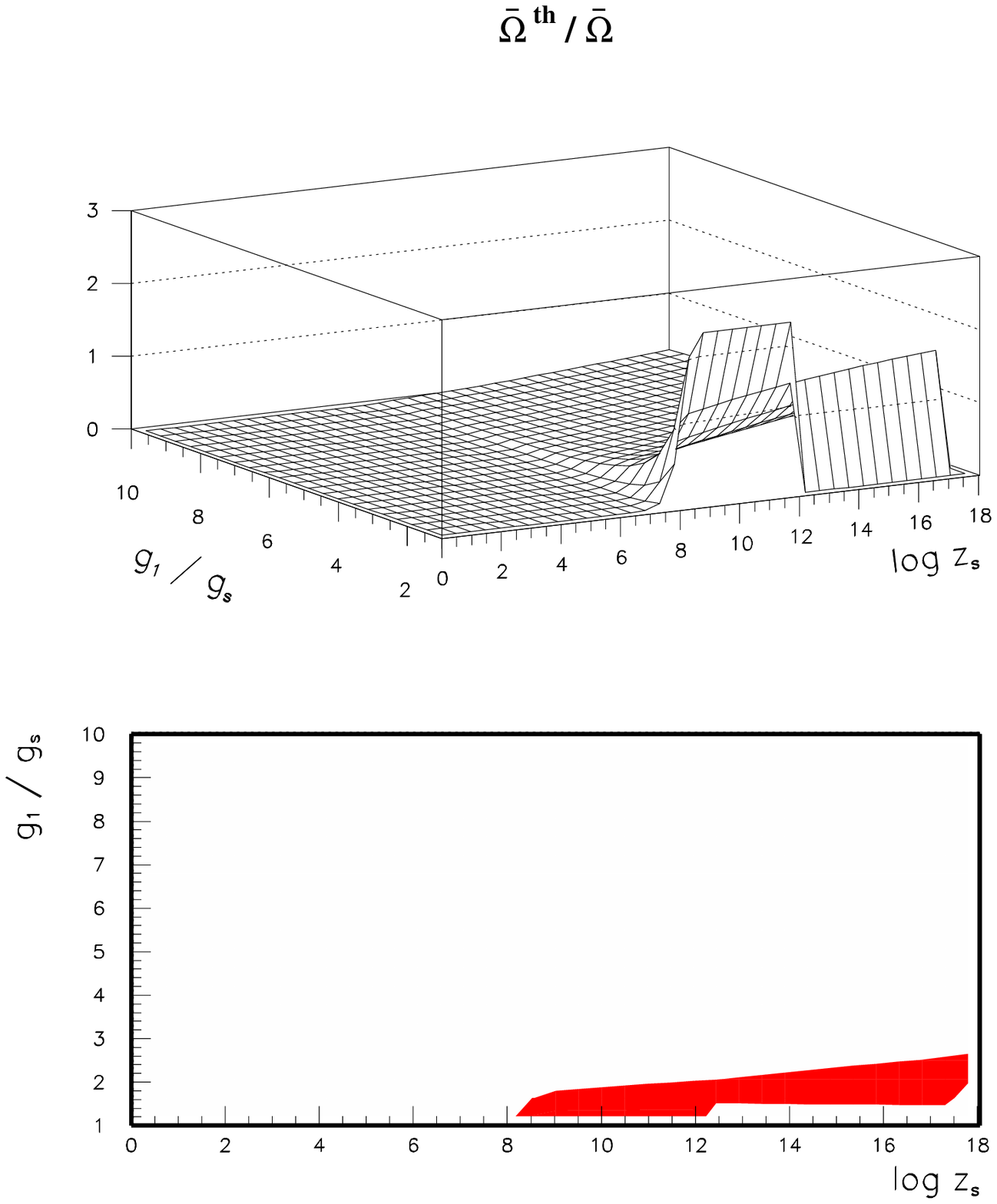}} \\
\end{tabular}
\end{center}
\vspace*{-1.5cm}
\caption[a]{We report the ratios 
$\overline{\Omega}^{\,{\rm max}}/\overline{\Omega}$ (left) and 
$\overline{\Omega}^{\,{\rm th}}/\overline{\Omega}$ (right)
as a function of $g_1/g_s$ and $\log{z_s}$  ($\overline{\Omega}$ 
is calculated for $T\,=\,1$ yr and SNR = 1). The lower contour plots 
show the regions where these ratios are greater than 1. 
The shaded area (bottom right) represents the region where the combination 
of the theoretical parameters is such that the corresponding 
$\overline{\Omega}^{\,{\rm th}}$ does not violate the BBN bound. 
As we can see the visibility region is reduced.
The difference between the shaded area in the right plot and the one 
in the left plot measures the error made by assuming as normalization of 
the spectrum not the theoretical one but the maximal one compatible 
with the BBN. The value $z_{s}\,=\,10^{8}$ roughly corresponds to 
$f_s\,\sim\,f_0$.
Notice that $\log$ denotes not the Neperian logarithm 
but the logarithm in ten basis.}
\label{minrat}
\end{figure}
After a transient period (whose precise duration will be fixed by the 
value of $g_1$), we will have that the radiation dominated evolution 
will take place when the value of the coupling constant will be of 
order one (i.e., $g_r\,\sim\,1$).

An interesting feature of this speculation is that the 
graviton spectra will not necessarily be monotonic \cite{gas1} 
(as the ones considered in the previous  analysis).
We then find interesting to apply our considerations also to this case.

The function $\omega (f)$ in the non-minimal model described 
above is given by \cite{gas1} \footnote{Notice that the form of $\omega(f)$ 
reported in \cite{gas1} differs from our expression only by 
logarithmic correction whose presence is, indeed, not relevant. We kept them 
only for sake of completeness.}
\begin{equation}
\omega (f)\,=\,
\left\{
\begin{array}{lc}
\dis \left(\,\frac{g_r}{g_1}\,\right)^{2/\sqrt{3}}\,
\left(\,\frac{f}{f_1}\,\right)^4\,
\left[\,\left(\,\frac{f_s}{f_1}\,\right)^{- \sigma}\,+\, 
\left(\,\frac{f_s}{f_1}\,\right)^{\sigma}\,\right]^2\,
\left(\,1\,-\,\ln{\frac{f_s}{f_1}}\,\right)^2
& \dis \qquad \qquad f_r\,<\,f\,\le\,f_s \,=\,
\frac{f_1}{z_s} \\ [15pt]
\dis \left(\,\frac{g_r}{g_1}\,\right)^{2/\sqrt{3}}\,
\left[\,\left(\,\frac{f}{f_1}\,\right)^{2 - \sigma}\,+\, 
\left(\,\frac{f}{f_1}\,\right)^{2 + \sigma}\,\right]^2\, 
\left(\,1\,-\,\ln{\frac{f}{f_1}}\,\right)^2 
& \dis \qquad \qquad f_s\,<\,f\,\le\,f_1 
\end{array}
\right.
\label{nonminth}
\end{equation}
where, in the present case 
\begin{equation}
f_1\,\simeq\,64.8\,\sqrt{g_1}\,\left(\,\frac{g_r}{g_1}\,
\right)^{1/2\sqrt{3}}\,\left(\,\frac{10^3}{n_r}\,\right)^{1/12}
\;{\rm GHz}\;, \qquad \qquad f_r\,=\,\left(\,\frac{g_r}{g_1}\,
\right)^{- 2/\sqrt{3}}\,f_1\;.
\end{equation}
The frequency $f_r$ corresponds to the 
onset of the radiation dominated evolution. 
If we adopt a purely phenomenological approach we can 
say that $f_r$ has to be bounded (from below) 
since we want the Universe to be 
radiation dominated not later than the BBN 
epoch. Thence, we have that $f_r\,>\,f_{\rm ns}$. Recalling 
the value of the nucleosynthesis frequency and 
assuming that $g_r\,\simeq\,1$ this condition 
implies $g_1\,\gaq\,8.2\,\times\,10^{-16}$. This simply 
means that in order not to conflict with 
the correct abundances of the light elements we have to 
require that the coupling constant should not be too small
when the curvature starts decreasing. Notice that 
for frequencies $f\,<\,f_r$ the spectrum evolves
as $f^{-3}$.
The ultra-violet branch of the spectrum is mainly 
originated by modes leaving the 
horizon during the stringy phase and re-entering when 
the dilaton coupling is still increasing.

Concerning the non-minimal spectra few comments are in order.
Owing to the fact that $g_1$ can be as small as 
$10^{-15}$ we have that the highest frequency of the spectrum 
can become substantially smaller than in the 
minimal case. Moreover, the spectrum might also be 
non-monotonic with a peak at $f_s$. Looking at the 
analytical form of the spectrum we see that 
this behavior occurs if $\sigma\,>\,2$.  
A non-monotonic logarithmic energy spectrum 
(with a maximum falling in the sensitivity region of 
the VIRGO pair) represents an interesting possibility. 

\begin{figure}[!hbp]
\begin{center}
\begin{tabular}{cc}
      \hbox{\epsfxsize = 8 cm  \epsffile{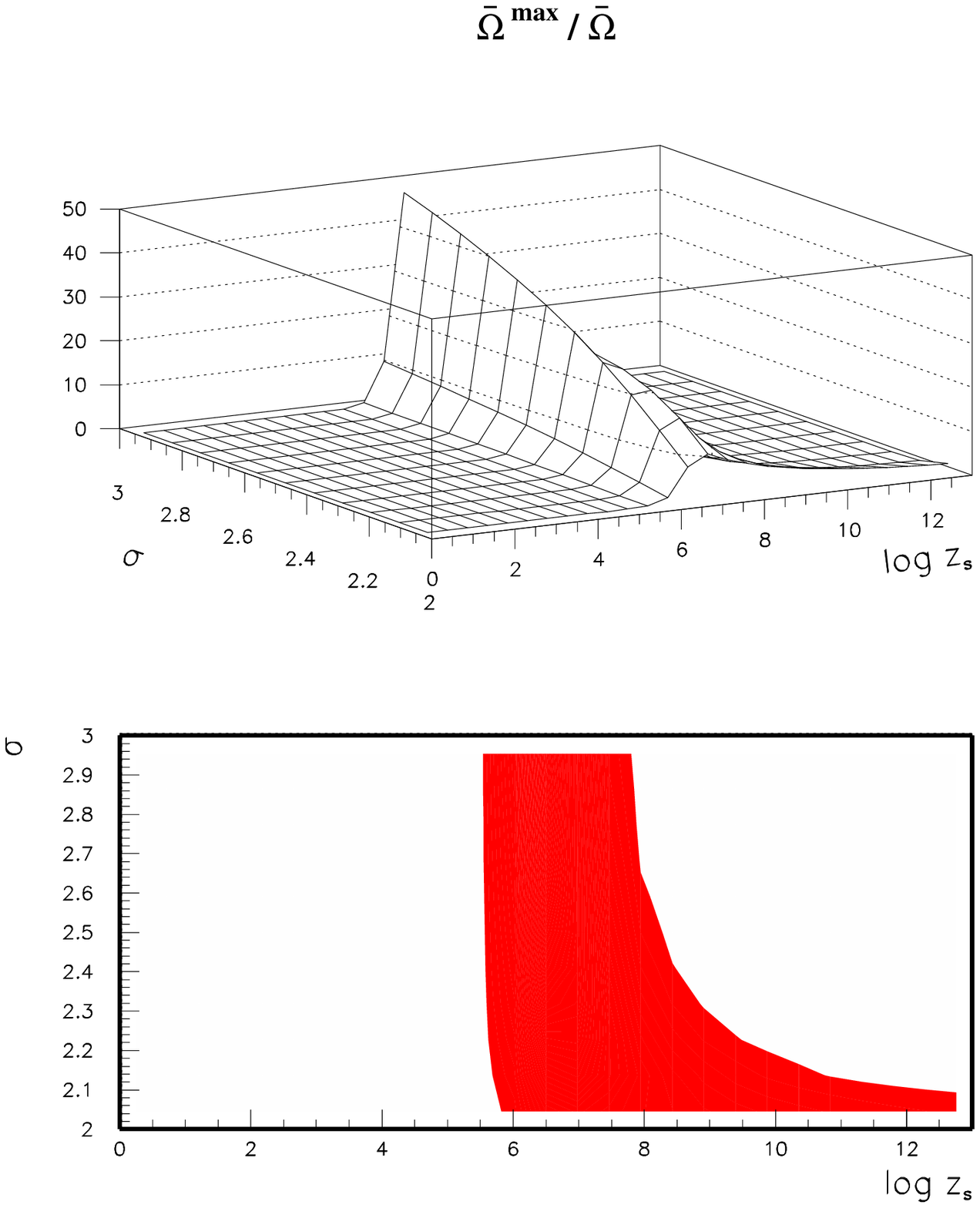}} &
      \hbox{\epsfxsize = 8 cm  \epsffile{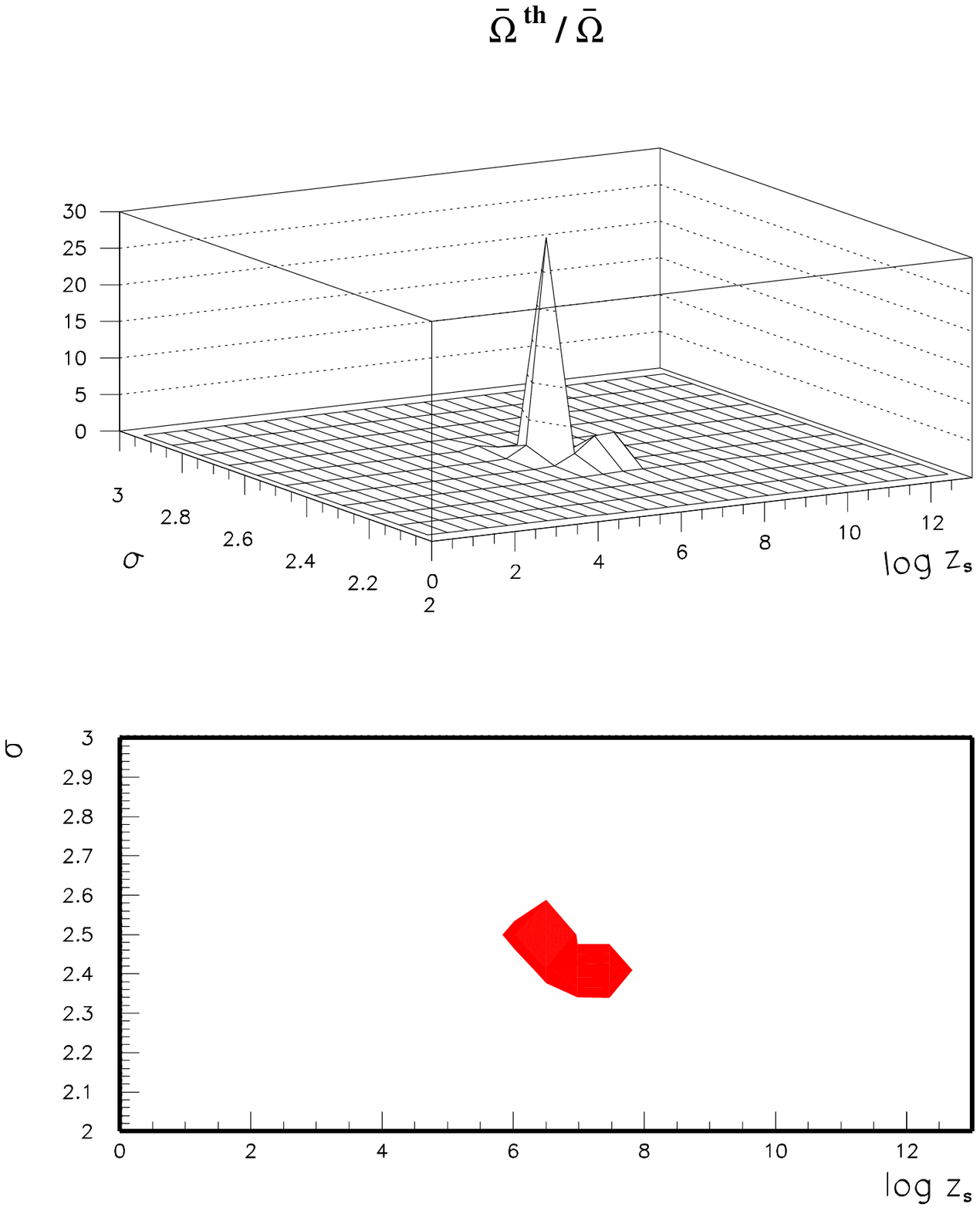}} \\
\end{tabular}
\end{center}
\vspace*{-1.5cm}
\caption[a]{In order to make clear the comparison with the 
visibility region of the minimal models, we report 
$\overline{\Omega}^{\,{\rm max}}/\overline{\Omega}$ (left) and
$\overline{\Omega}^{\,{\rm th}}/\overline{\Omega}$ (right)
as a function of $\sigma$ and of the $\log{z_s}$ in the non-minimal 
scenario. Notice that we took $g_1\,=\,10^{-12}$,  $n_r\,=\,10^3$, 
and $g_r\,=\,1$. As for Fig. \ref{minrat}, the shaded areas in the 
lower contour plots represent the region where each ratio is greater 
than 1, and, in the case of the right plot, also the BBN is satisfied.}
\label{nminrat}
\end{figure}

The results of our calculation for $g_1\,=\,10^{-12}$,
 $n_r\,=\,10^3$, $g_r\,=\,1$,and $\sigma\,>\,2$ are reported 
in Fig. \ref{nminrat}. As done in the case of minimal spectra 
we analyse the visibility window in the plane of the relevant 
parameters of the model. As we can see from the  left part of 
Fig. \ref{nminrat} the region compatible with the BBN is rather 
large but it shrinks when we impose the theoretical normalization 
( right part of Fig. \ref{nminrat}) which is always 
smaller than the maximal normalization allowed by BBN. 
It is interesting to compare directly the three dimensional plots appearing 
in Fig. \ref{minrat} with the corresponding three dimensional plots 
of Fig. \ref{nminrat}. We can see that the regions of parameter space 
where $\overline{\Omega}^{\,\rm max}/\overline{\Omega}$ and  
$\overline{\Omega}^{\,\rm th}/\overline{\Omega}$ are larger than one 
is larger in the case of minimal models. However, the shaded region in
the case of minimal models corresponds to ratios 
$\overline{\Omega}^{\,\rm max}/\overline{\Omega}$ and 
$\overline{\Omega}^{\,\rm th}/\overline{\Omega}$ which can be 
 3 or 2, respectively. On the other hand the shaded region in 
the case of Fig. \ref{nminrat} corresponds to ratios 
$\overline{\Omega}^{\rm max}/\overline{\Omega}$ and 
$\overline{\Omega}^{\rm th}/\overline{\Omega}$ which can be, 
respectively, as large as 50 or 25. So, in the latter case the 
signal is larger for a smaller region of the parameter space. 

\begin{figure}[!ht]
\begin{center}
\begin{tabular}{cc}
      \hbox{\epsfxsize = 8 cm  \epsffile{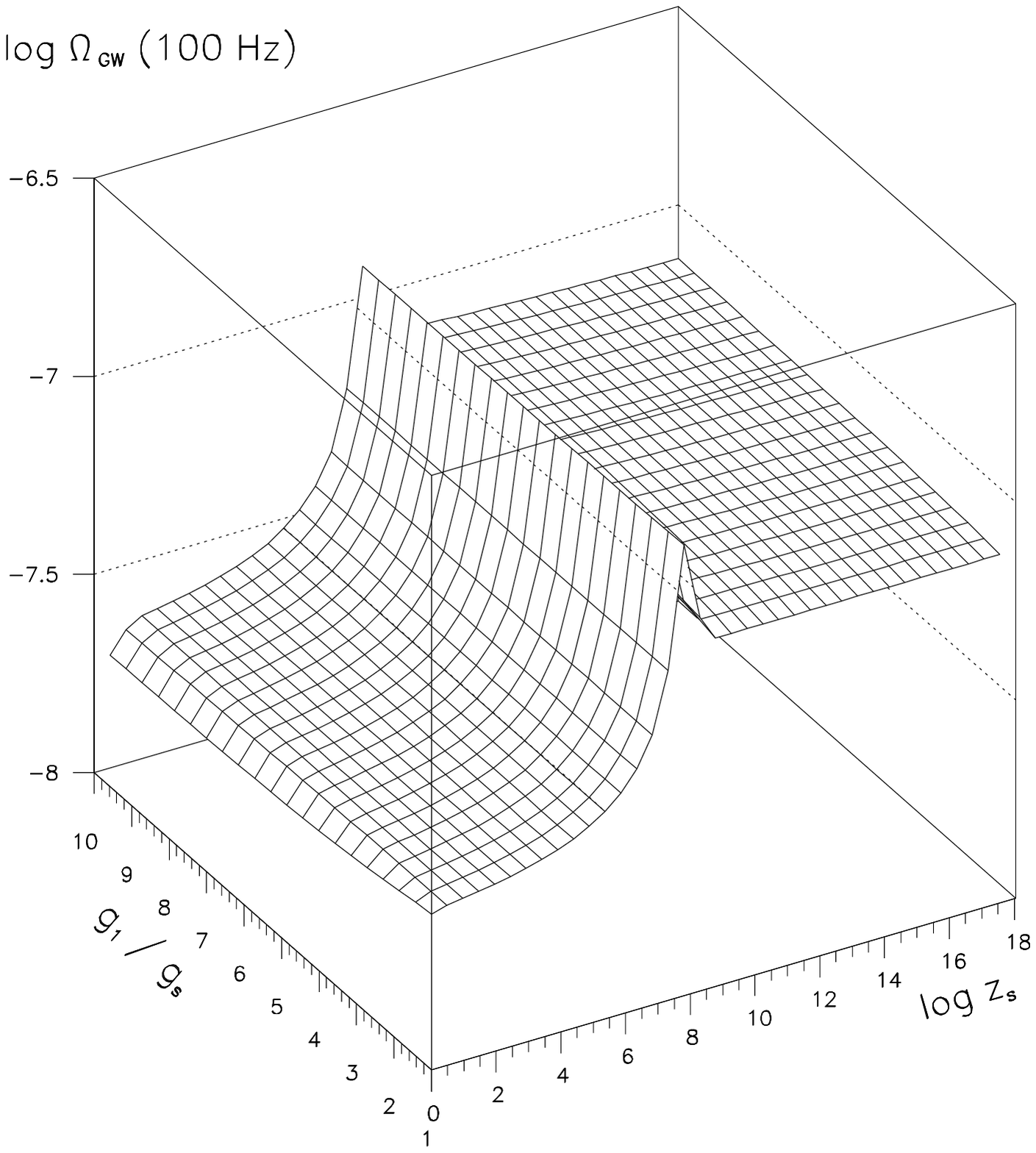}} &
      \hbox{\epsfxsize = 8 cm  \epsffile{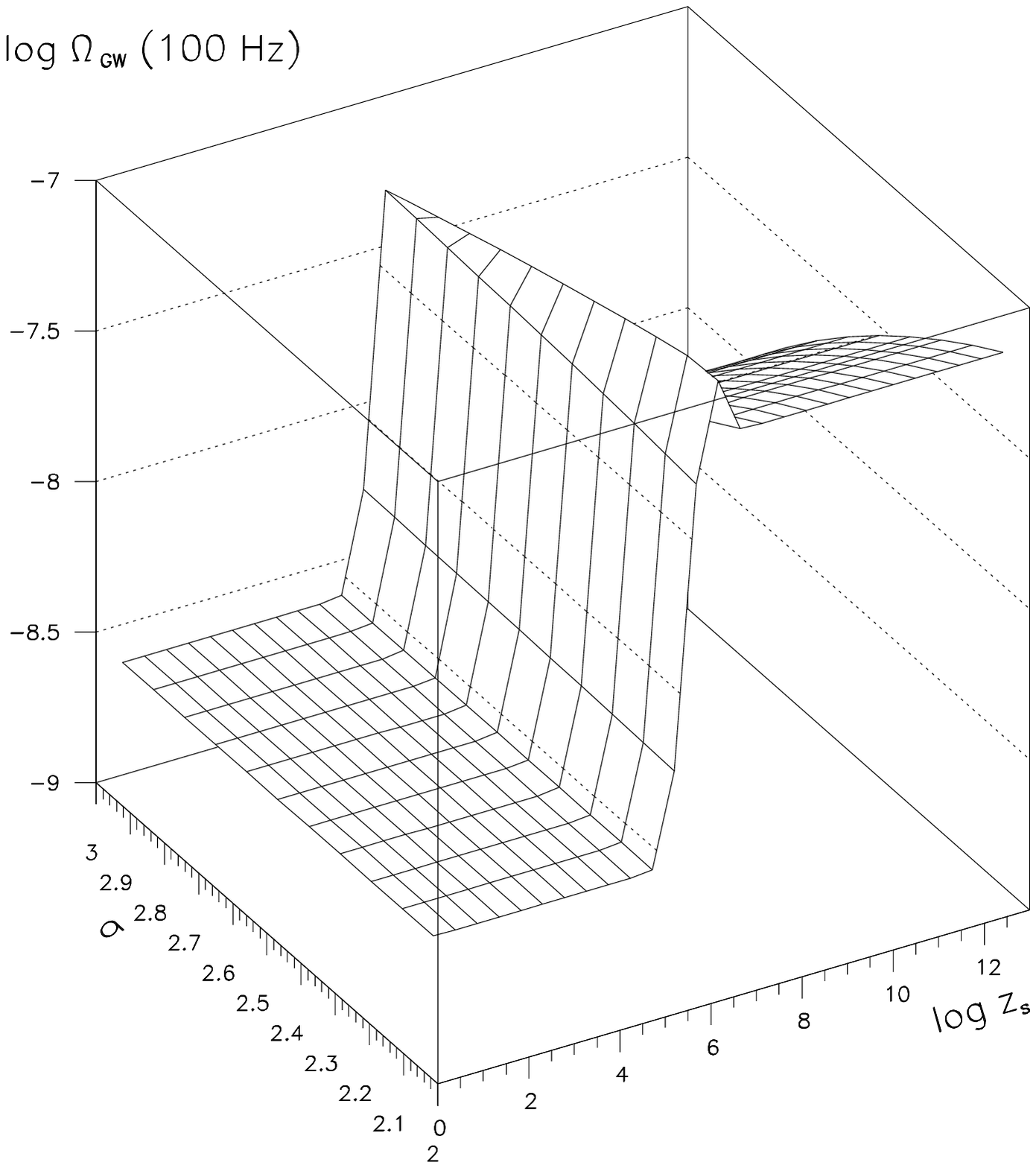}} \\
\end{tabular}
\end{center}
\vspace*{-1.0cm}
\caption[a]{We report the logarithm of the sensitivity of the 
VIRGO pair at 100 Hz for $T\,=\,1$ yr and SNR = 1 in the case of 
minimal (left plot) and non-minimal (right plot) energy spectra.}
\label{figsen}
\end{figure}

As we stressed in the previous Section, $\overline{\Omega}$ represents  
the sensitivity of the VIRGO pair to a given spectrum whose functional 
form is given by $\omega(f)$. One might be interested, in principle, 
in the sensitivity of the VIRGO pair at a specific frequency $f^*$. 
This can be easily computed by multiplying $\overline{\Omega}$ by 
$\omega (f^*)$. 
In Fig. \ref{figsen} we show the sensitivity of the VIRGO pair at 
the frequency $f^*\,=\,100$ Hz, both, for the minimal and non-minimal 
models  considered in the present Section. One can easily discuss the 
same quantity for any other frequency in the operating window of the 
VIRGO detectors.

\renewcommand{\theequation}{3.\arabic{equation}}
\setcounter{equation}{0}
\section{Noise reduction and the visibility region of a VIRGO pair}

There are two ways of looking at the calculations reported in this 
paper. One can look at these ideas from a purely theoretical 
perspective. In this respect we presented a study of the 
sensitivity of a pair of VIRGO detectors to string cosmological 
gravitons. There is also a second way of looking at our exercise. 
Let us take at face value the results we obtained and let us ask 
in what way we can enlarge the visibility region of the VIRGO pair. 
In this type of approach the specific form of graviton spectrum is 
not strictly essential. We could use, in principle, any motivated 
theoretical spectrum. As we stressed, we will use string 
cosmological spectra because, on one hand, they are theoretically 
motivated and, on the other hand, they give us a signal which
could be, in principle detected. Of course, there are other well 
motivated spectra (like the ones provided by ordinary inflationary 
models). However, the signal would be, to begin with, quite small. 

In this Section we will then consider the following problem. 
Given a pair of VIRGO detectors, we suppose to be able, 
by some means, to reduce, in a selective fashion, 
the contribution of a specific noise source to the detectors
output. The question we ought to address is how the visibility 
regions will be modified with respect to the case in which the 
selective noise reduction is not present. 
We will study the problem for the pair of VIRGO detectors 
considered in the previous Sections, i.e., for identical detectors 
with NPS given in Eq. (\ref{NPS}), and characterized by the 
overlap reduction function of the case A of Fig. \ref{over}.
As for the theoretical graviton spectrum we will focus our 
attention on the case of minimal models considered in Section III.A, 
 with the same parameters used to produce Fig. \ref{minrat}. 
Also here, the quantity $\overline{\Omega}$ will be computed 
for $T\,=\,1$ yr and SNR = 1.
 
\begin{figure}[!t]
\centerline{\epsfxsize = 7 cm  \epsffile{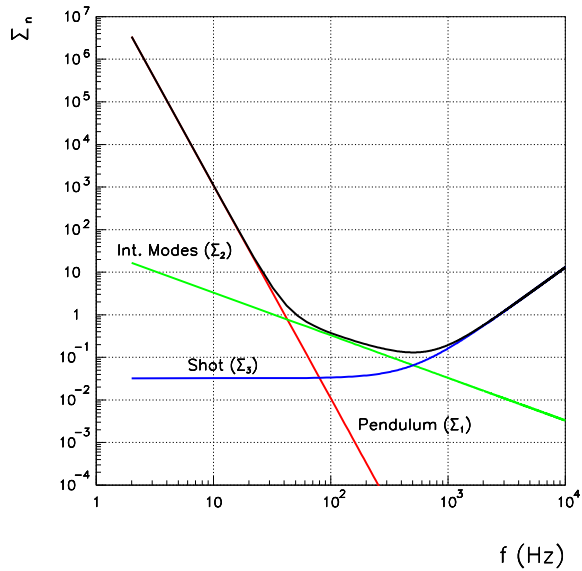}} 
\vspace*{-1.0cm}
\caption[a]{The analytical fit of the rescaled noise power 
spectrum $\Sigma_n$ defined in Eq. (\ref{NPS}) in the case 
of the VIRGO detector. With the full (thick) line we denote 
the total NPS. We also report the separated contribution of the three
main (Gaussian and stationary) sources of noise.}
\label{noise}
\end{figure}

As shown in Section II the NPS is characterized by three 
dimension-less numbers $\Sigma_{1,2,3}$, and two frequencies 
$f_a$ and $f_{b}$. Roughly, $\Sigma_1$ and $\Sigma_2$ 
control, respectively, the strength of the  pendulum 
and pendulum's internal modes noise, whereas $\Sigma_3$ is 
related to the shot noise  (see Ref. \cite{sau} for an 
accurate description of the phenomena responsible of these 
noises). Below the frequency $f_b$ the NPS is dominated by 
the seismic noise (assumed to be infinitum in Eq. (\ref{NPS})). 
The frequency $f_a$ is, roughly, where the NPS gets its minimum. 
The frequency behavior of this three contributions and of the 
total NPS is shown in Fig. \ref{noise}. The stochastic processes 
associated with each source of noise are assumed to be Gaussian 
and stationary. 

In the following, without entering in details concerning the actual 
experimental  strategy adopted for the noise reduction, we will 
suppose to be able to reduce each of the coefficients $\Sigma_i$ by 
keeping the other fixed. In order to make our notation simpler we 
define a ``reduction vector''
\begin{equation}
\vec{\rho}\,=\,(\rho_1, \rho_2, \rho_3)\;,
\end{equation}
whose components define the reduction, respectively, of the seismic, 
thermal and shot noises with respect to their fiducial values 
appearing in Eq. (\ref{NPS})  (corresponding to the case 
$\vec{\rho}\,=\,(1, 1, 1)$).

\begin{figure}[!ht]
\begin{center}
\begin{tabular}{cc}
      \hbox{\epsfxsize = 8 cm  \epsffile{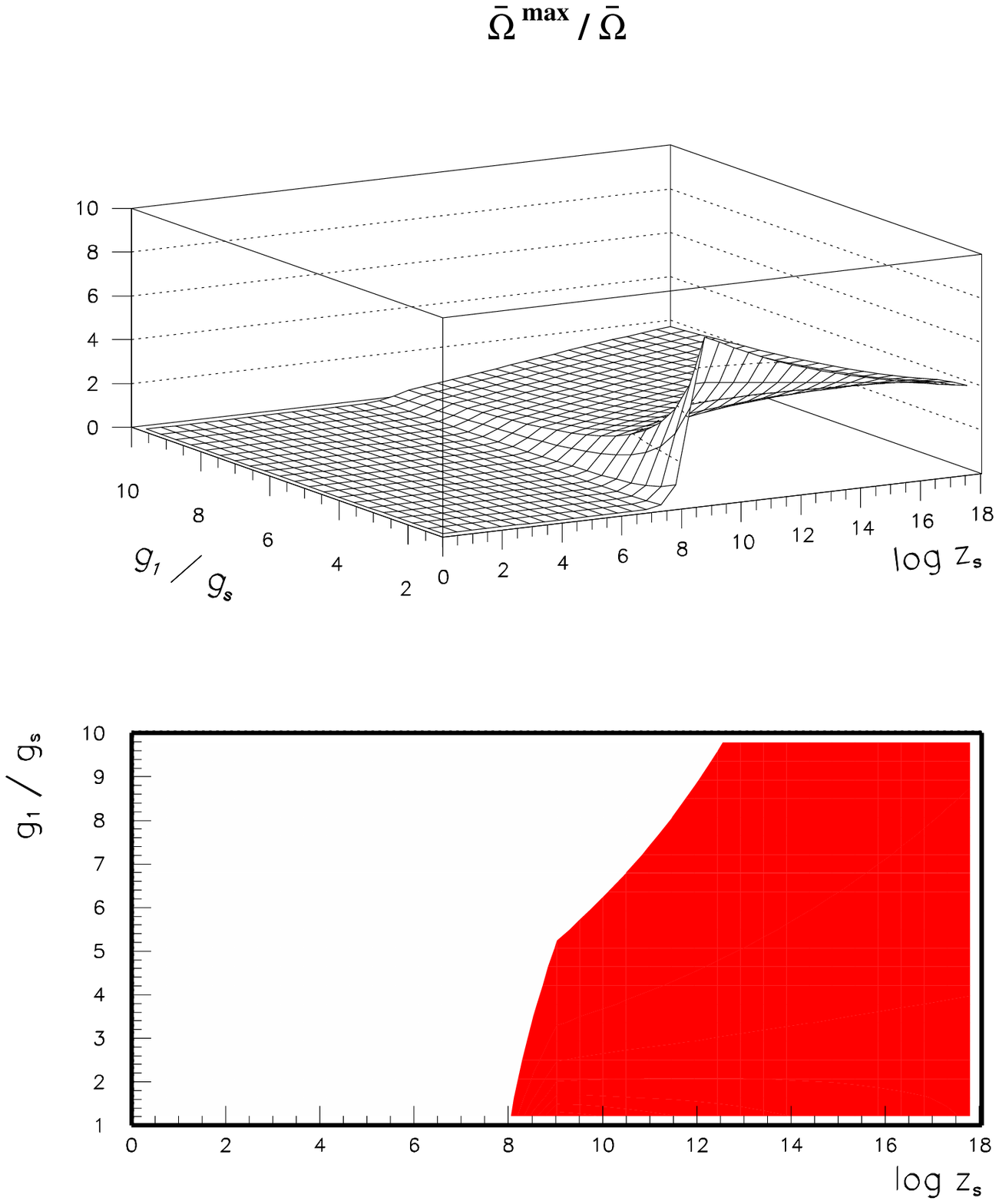}} &
      \hbox{\epsfxsize = 8 cm  \epsffile{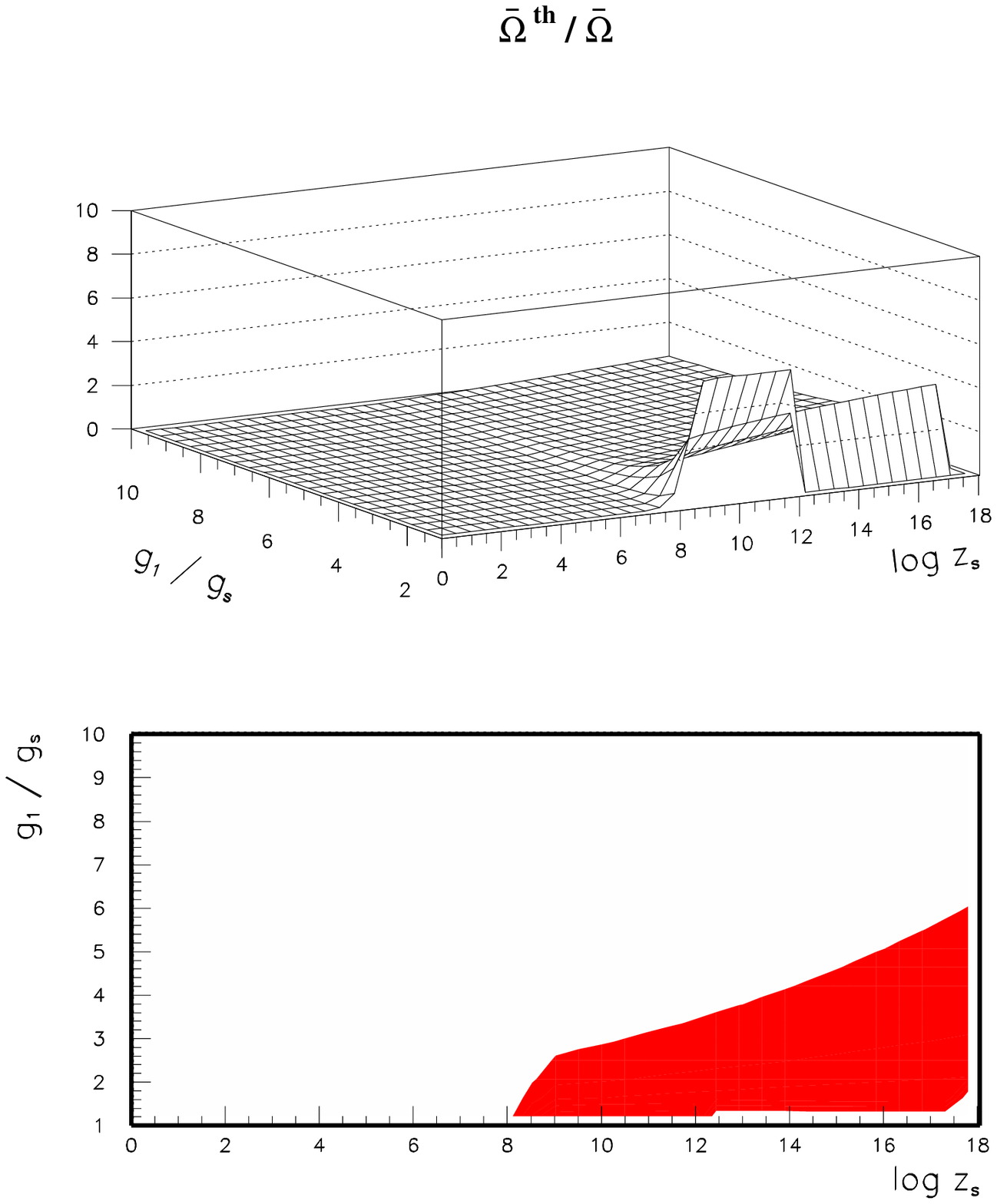}} \\
\end{tabular}
\end{center}
\vspace*{-1.5cm}
\caption[a]{We report the ratios 
$\overline{\Omega}^{\,{\rm max}}/\overline{\Omega}$ (left plots), and 
$\overline{\Omega}^{\,{\rm th}}/\overline{\Omega}$ (right plots) 
in the case in which the shot noise and the noise related to the 
pendulum's internal modes are not reduced, whereas the pendulum noise 
is diminished by a factor of ten with respect to the values quoted in 
Eq. (\ref{NPS}), i.e., $\vec{\rho}\,=\,(0.1, 1, 1)$.}
\label{noired1}
\end{figure}

As shown in Fig. \ref{noise} the pendulum noise dominates the 
sensitivity of the detectors in the low frequency region, namely 
below about 40 Hz. In Fig. \ref{noired1} we report the results of 
our calculation for the case $\vec{\rho}\,=\,(0.1, 1, 1)$. 
Here the parameters of the theoretical spectrum are exactly the 
same as in Fig. \ref{minrat}. The only change is given by a reduction 
of the pendulum noise. From the comparison between Fig. \ref{noired1} 
and Fig. \ref{minrat}, we see that the visibility region in the 
parameter space of our model gets immediately larger especially 
towards the region of small $g_s$. This enlargement is quite 
interesting especially in terms of 
$\overline{\Omega}^{\,{\rm th}}/\overline{\Omega}$.

\begin{figure}[!hb]
\begin{center}
\begin{tabular}{cc}
      \hbox{\epsfxsize = 8 cm  \epsffile{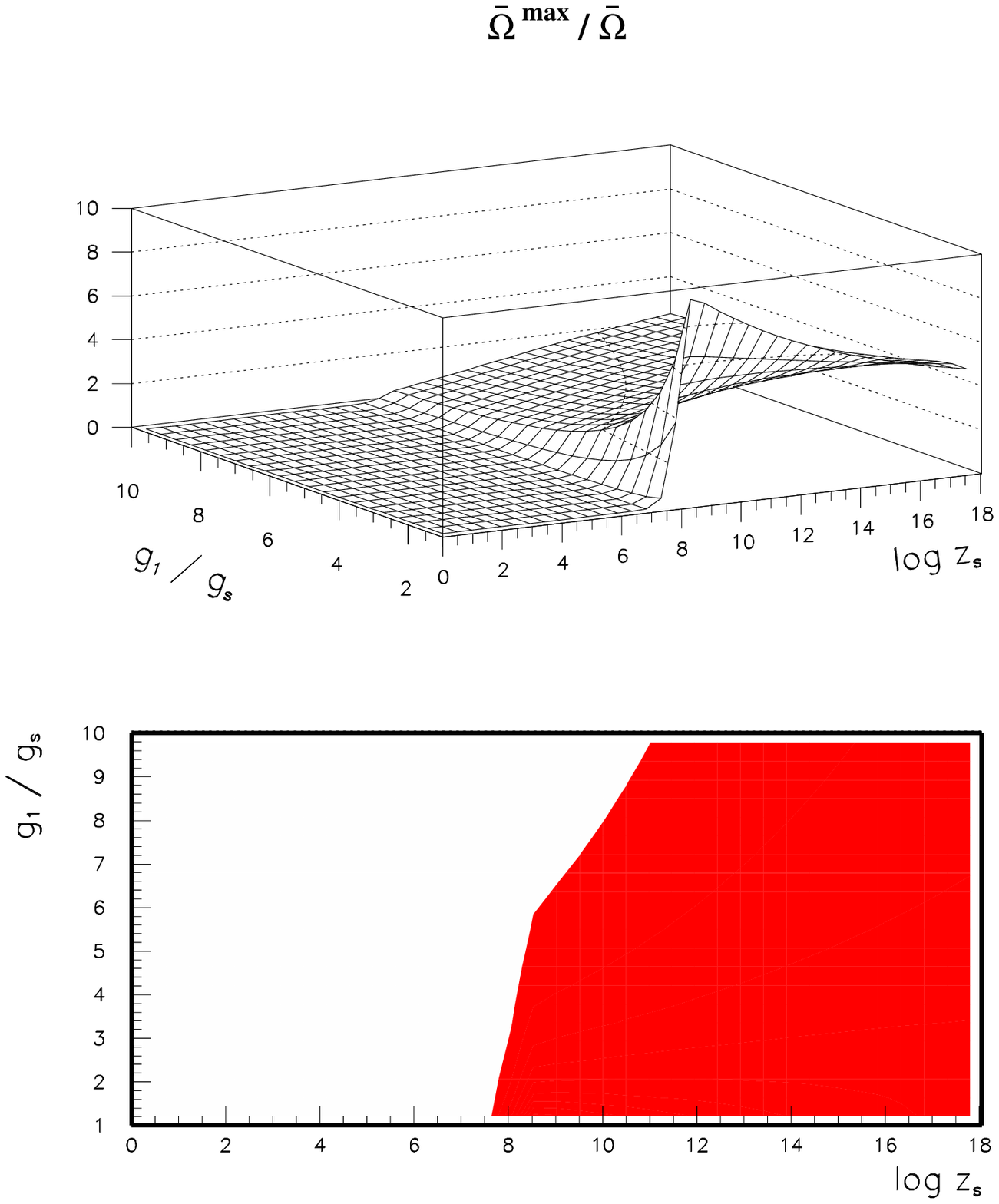}} &
      \hbox{\epsfxsize = 8 cm  \epsffile{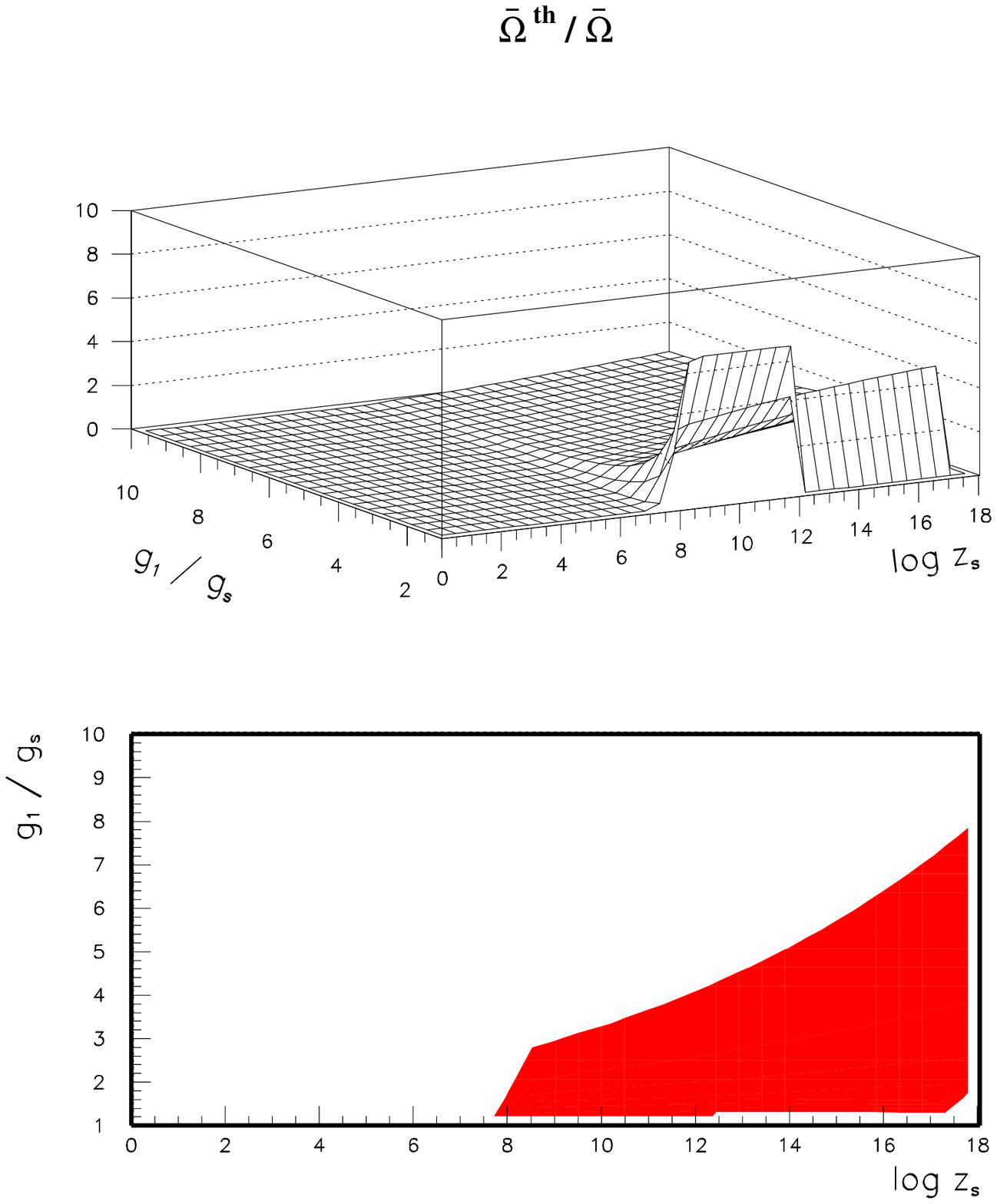}} \\
\end{tabular}
\end{center}
\vspace*{-1.5cm}
\caption[a]{We report the result of selective reduction in the case 
where the  noise cause by the pendulum's internal modes is reduced 
by a factor of ten, whereas the  pendulum and shot contributions 
are left unchanged, i.e., $\vec{\rho}\,=\,(1, 0.1, 1)$.}
\label{noired2}
\end{figure}

\begin{figure}[!ht]
\begin{center}
\begin{tabular}{cc}
      \hbox{\epsfxsize = 8 cm  \epsffile{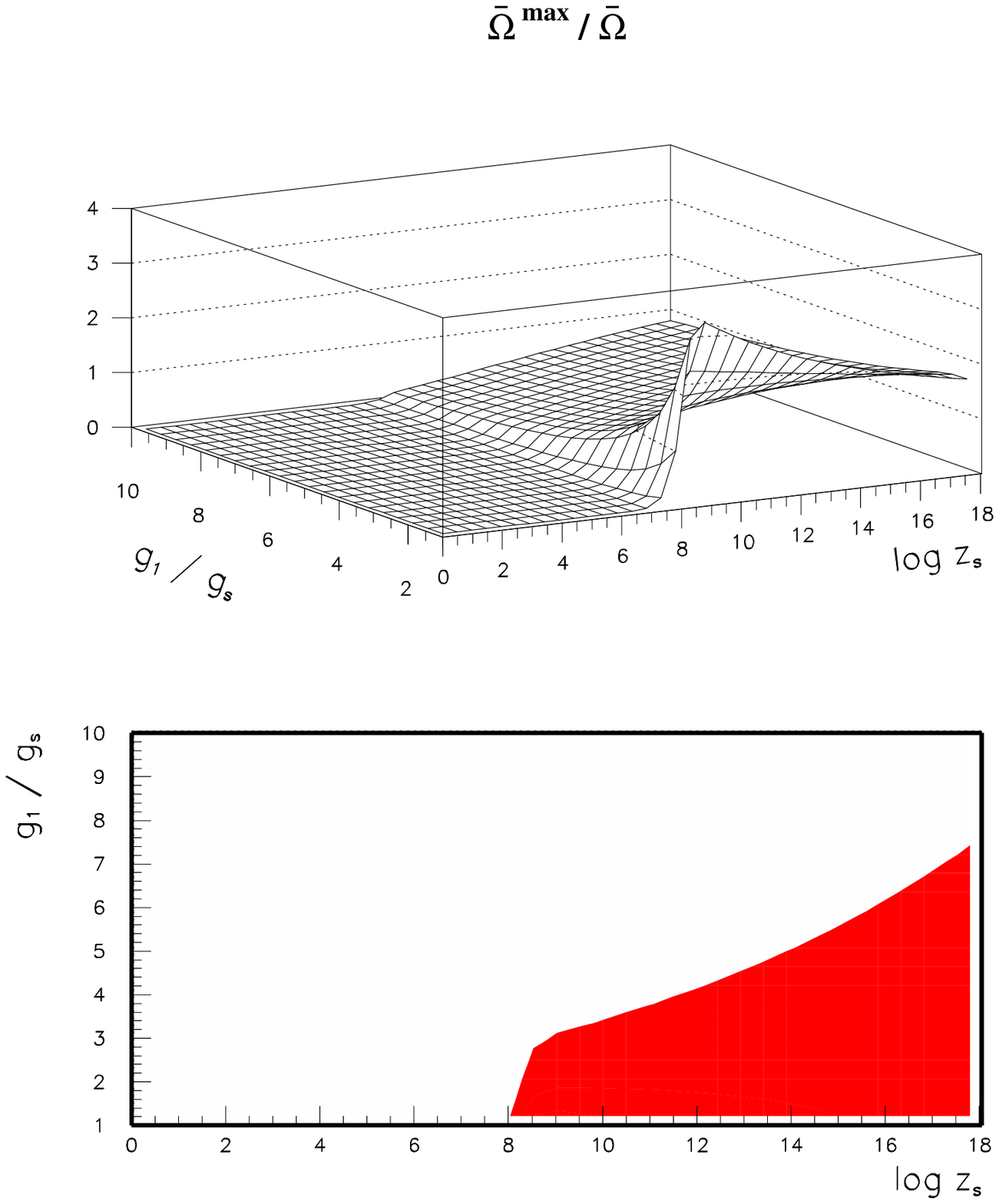}} &
      \hbox{\epsfxsize = 8 cm  \epsffile{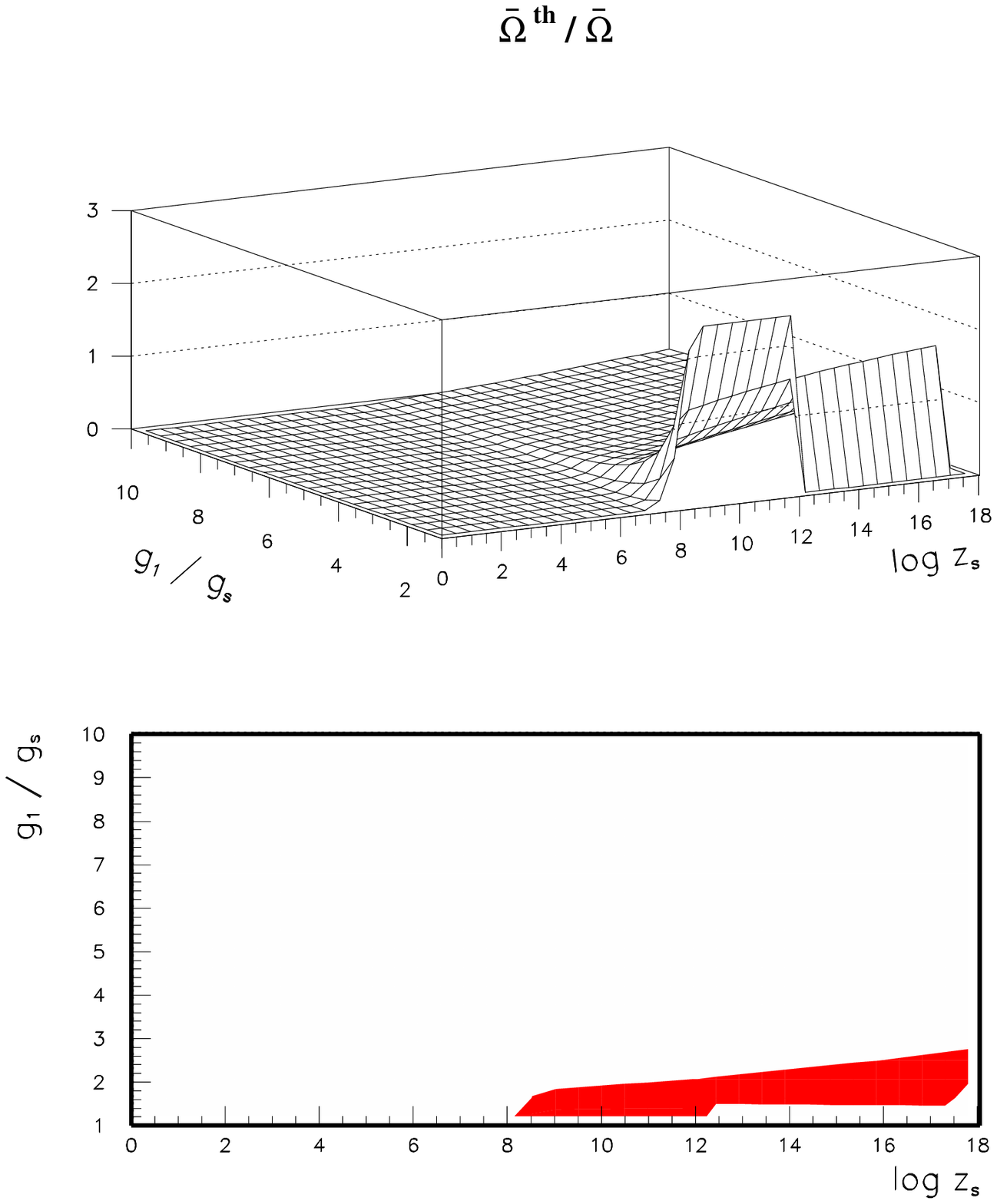}} \\
\end{tabular}
\end{center}
\vspace*{-1.5cm}
\caption[a]{We report the same quantities discussed in Fig. \ref{noired1} 
for the case $\vec{\rho}\,=\,(1, 1, 0.1)$. Thhe shaded areas in the lower 
plots are the relevant visibility regions which should be compared 
with the shaded regions in the lower plots of Figs. \ref{noired1} and 
\ref{noired2}. By direct comparison we can argue that a reduction in the 
shot noise (by a factor of ten) is not as efficient as a reduction, by 
the same amount, in the thermal noise components.}
\label{noired3}
\end{figure}

In the frequency region between 50 and 500 Hz the performances of the 
detectors are, essentially, limited by the  pendulum's internal 
modes noise. The results obtained for a selective reduction of this 
component are summarized in Fig. \ref{noired2},  where the 
pendulum and shot noises are left unchanged but the internal modes
component is reduced by a factor of ten (in Fig. \ref{noired3}). As we 
can see the visibility region gets larger and the increase in the area 
is comparable with the one obtained by selecting only the  pendulum 
noise. 

\begin{figure}[!ht]
\begin{center}
\begin{tabular}{cc}
      \hbox{\epsfxsize = 8 cm  \epsffile{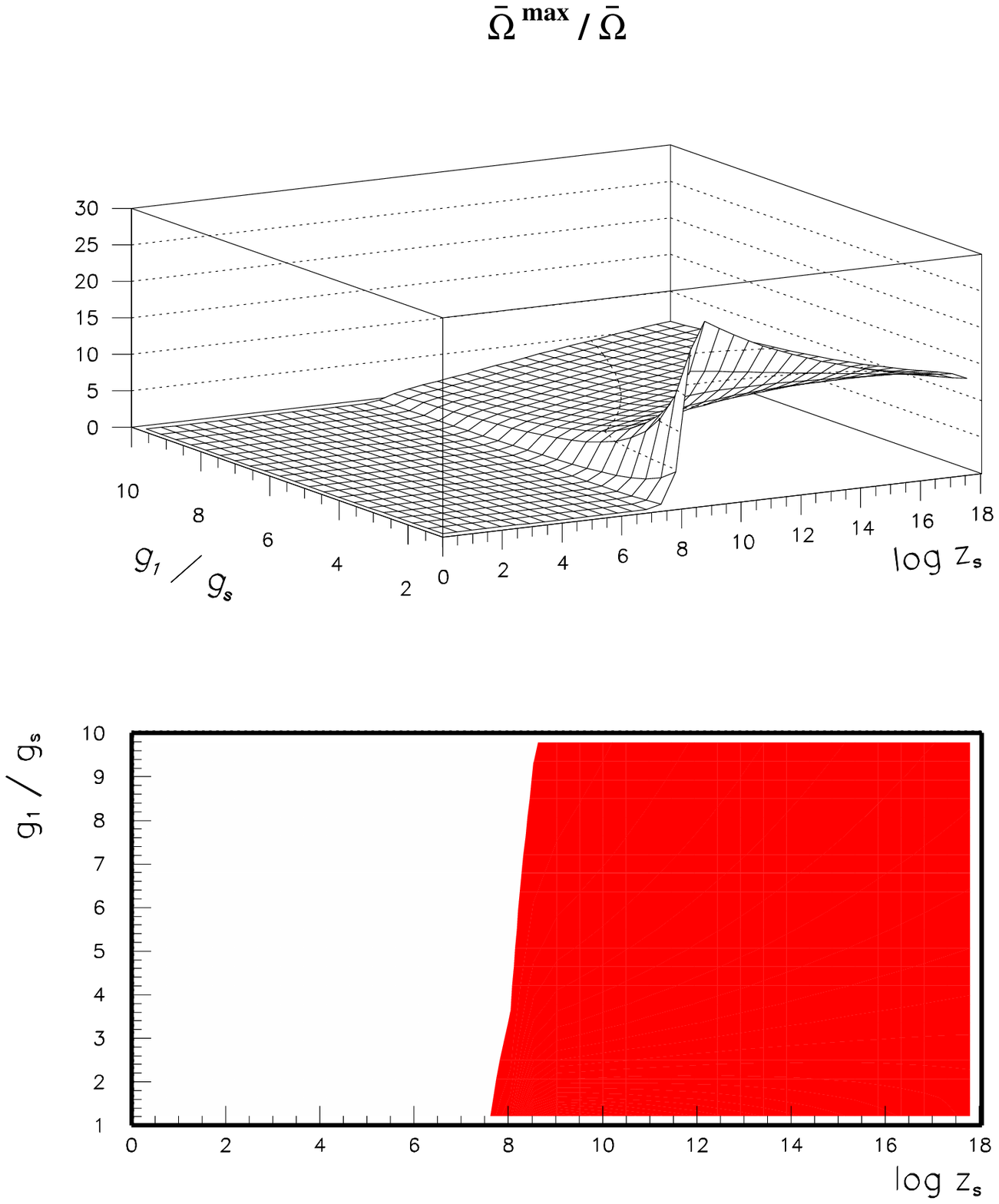}} &
      \hbox{\epsfxsize = 8 cm  \epsffile{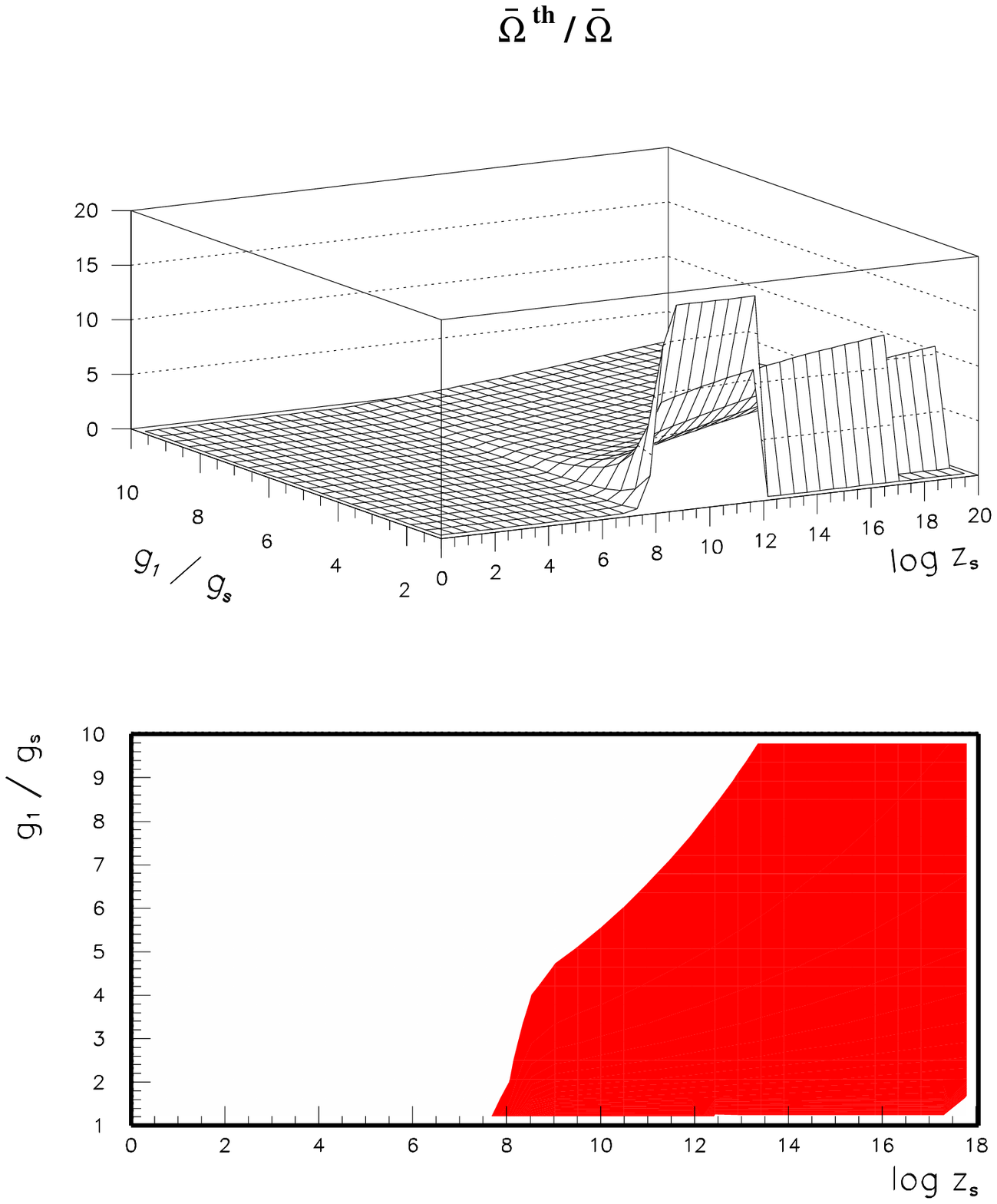}} \\
\end{tabular}
\end{center}
\vspace*{-1.5cm}
\caption[a]{ We illustrate the case of a 
simultaneous reduction of $\Sigma_1$ and $\Sigma_2$ by a factor 10, 
whereas $\Sigma_3$ is the same of Eq. (\ref{NPS}), i.e., 
$\vec{\rho}\,=\,(0.1, 0.1, 1)$.}
\label{noired12}
\end{figure}

Finally, for sake of completeness, we want to discuss the case of the 
shot noise, i.e., the noise characteristic of the detector above 500 
Hz. Our results for $\rho_3\,=\,0.1$ are reported in Fig. \ref{noired12}.
As we can see by comparing Figs. \ref{noired1}, \ref{noired2}, and 
\ref{noired3} we gain much more in visibility by reducing the thermal 
noise  components than by reducing the shot noise. 
In Fig. \ref{noired3} the shot noise is reduced by one tenth but the 
visibility region does not increase by much (left plot). This result 
is consequence of the fact that, as shown by Fig. \ref{noise}, the 
shot noise contribution to the NPS starts to be relevant for 
$f\,\sim\,1$ kHz, i.e., in a frequency region where the overlap between 
the detectors begins to deteriorate (see Fig. \ref{over}). 
In Figs. \ref{noired1} and \ref{noired2} the thermal noise is reduced 
by one tenth and the increase in the visibility region is, comparatively, 
larger. This shows, amusingly enough, that a reduction in the shot noise 
will lead to an effect whose practical relevance is already questionable 
at the level of our analysis. Notice that a selective noise reduction 
can be also discussed in the case of a purely flat spectrum \cite{bg}.

In order to conclude this Section we want to show the combined action of
the simultaneous reduction of  both the components of the thermal 
noise. In Fig. \ref{noired12}, owing to the results of our analysis we 
kept the shot noise fixed but we reduced both the thermal and seismic 
noises by a factor of ten. Clearly we observe a consistent increase 
in the visibility region.
However, even if a combined reduction of these components 
cannot be achieved we want to stress that already a reduction of 
the pendulum's internal modes noise alone (by one tenth) 
can be of relevant practical interest.

\renewcommand{\theequation}{5.\arabic{equation}}
\setcounter{equation}{0}
\section{Discussion and executive summary}

There are no compelling reasons why one should not consider 
the appealing theoretical possibility of a second VIRGO detector
coaligned with the first one. Moreover, recent experimental 
suggestions seem coherently directed towards this goal \cite{gia}. 
While the location of the second detector is still under
debate we presented a theoretical analysis of some of the 
scientific opportunities suggested by this proposal.

We focused our attention on possible cosmological sources 
of relic gravitons and we limited our attention to the case 
of stochastic and isotropic background produced by the adiabatic
variation of the backgound geometry. In the framework of these 
models we can certainly argue that in order to have a large 
signal in the frequency window covered by VIRGO we have to focus 
our attention on models where the logarithmic energy spectrum
increases at large frequencies. Alternatively we have to look 
for models where the logarithmic energy spectrum exhibits some 
bump in the vicinity of the VIRGO  operating window. 
If the logarithmic energy spectra are decreasing as a function 
of the present frequency (as it happens in ordinary inflationary 
models) the large scale (CMB) constraints forbid a large signal 
at high frequencies. In the case of string cosmological models 
the situation seems more rosy and, therefore, we use these models 
as a theoretical laboratory in order to investigate, in a 
specific model the possible improvements of a possible VIRGO pair. 
The choice of a specific model is, in some sense, mandatory. In 
fact, owing to the form of the SNR we can immediately see that 
different models lead to different SNR not only because the amplitude 
of the signal differs in different models. Indeed, one can convince 
himself that two models with the same amplitude at $100$ Hz but different 
spectral behaviors between 2 Hz and 10 kHz lead to different SNR.

In order to analyze the sensitivity of the VIRGO pair we  described 
a semi-analytical technique whose main advantage is to produce the 
sensitivity of the VIRGO pair to a theoretical spectrum of arbitrary 
slopes and amplitudes. The theoretical error is estimated, in our 
approach, by requiring the compatibility with all the phenomenological 
bounds applicable to the graviton spectra. As an intersting example,
we asked what is the sensitivity of a VIRGO pair to string cosmological 
spectra {\em assuming} that a second VIRGO detector (coaligned with 
the first one) is built in a european site. By assuming that the second 
VIRGO detector has the same features of the first one we computed the 
SNR and the related sensitivity achievable after one year of observation 
in the case of string cosmological spectra. 

By using the string cosmological spectra as a theoretical laboratory 
we then studied some possible noise reduction. Our main goal, in this 
respect, has been to spot what kind of stationary and stochastic noise 
should be reduced in order to increase the visibility region of the 
VIRGO pair in the parameter space of the theoretical models under 
considerations. Our main result is that a selective reduction of each 
of the three main sources of noise is not equivalent. A reduction in 
the shot noise by a factor of ten does not increase significantly the 
visibility region of the VIRGO pair. A selective reduction of 
the thermal noise components is far more efficient. In particular, 
we could see that a reduction (of one tenth) of the 
pendulum's internal modes increases the visibility region of four 
times. The simultaneous reduction of the  two components of the 
thermal noise leads to an even more relevant increase. 

The construction of a second VIRGO detector coaligned with the first 
one and an overall reduction of the thermal noise of each detector 
of the pair leads to what we called ``upgraded VIRGO'' program. The 
results presented in this paper are obtained in the case of a 
particularly promising class of theoretical models but can be generally 
applied to any logarithmic energy spectrum with similar qualitative 
results. However, owing to the non-linearities present in the evaluation 
of the SNR it would not be correct assess that they hold, quantitatively, 
without change. We hope that our results and our suggestions may turn out 
to be useful in the actual process of design of the upgraded VIRGO program 
\cite{gia}.

\section*{Acknowledgements}
We would like to thank A. Giazotto for very 
useful hints and for his kind interest in this investigation.

\newpage
\begin{appendix}

\renewcommand{\theequation}{B.\arabic{equation}}
\setcounter{equation}{0}
\section{ The normalization of the overlap reduction function}

In this Appendix we discuss the reduction in sensitivity due to the fact 
that, in general, these detectors will not be either coincident 
or coaligned. This effect is quantified by the (dimensionless) 
overlap reduction function $\gamma (f)$ appearing in Eq. (\ref{SNR}). 
Suppose that we have a gravitational wave propagating along 
a generic direction characterized, in spherical coordinates, 
by the unit vector $ \hat{\Omega} = (\cos{\phi}\,\sin{\theta},\,
\sin{\phi}\,\sin{\theta},\,\cos{\theta})$. If we now introduce a pair 
of orthogonal unit vectors directed in the plane perpendicular to 
$\hat{\Omega}$
\beq\label{vers}
\hat{m} (\hat{\Omega}) \equiv (\cos{\phi}\,\cos{\theta},\,
\sin{\phi}\,\cos{\theta},\, - \sin{\theta}) \qquad \qquad
\hat{n} (\hat{\Omega}) \equiv (\sin{\phi},\,- \cos{\phi},\,0)\,,
\eeq
the polarization tensors can be written, in terms of the polarization angle 
$\psi$ of the GW, as
\begin{eqnarray}
\varepsilon^+ (\hat{\Omega},\psi) &=& e^+ (\hat{\Omega})
\cos{2 \psi}\,-\,e^\times (\hat{\Omega}) \sin{2 \psi} \nonumber \\
\varepsilon^\times (\hat{\Omega},\psi) &=& e^+ (\hat{\Omega})
\sin{2 \psi}\,+\,e^\times (\hat{\Omega}) \cos{2 \psi}\,,
\end{eqnarray}
where 
\beq\label{polten}
e^+(\hat{\Omega}) = \hat{m} (\hat{\Omega}) \otimes 
\hat{m} (\hat{\Omega}) - \hat{n} (\hat{\Omega}) \otimes 
\hat{n} (\hat{\Omega}) \qquad 
e^{\times}(\hat{\Omega}) = \hat{m} (\hat{\Omega}) \otimes 
\hat{n} (\hat{\Omega}) + \hat{n} (\hat{\Omega}) \otimes 
\hat{m} (\hat{\Omega})
\eeq
with the normalization
\bdis
\mbox{Tr}\,\{ e^A(\hat{\Omega})\,e^{A'}(\hat{\Omega}) \} = 
2\,\delta^{AA'}\,.
\edis
If the graviton background is isotropic and 
unpolarized we will have that 
\beq\label{gamma}
\gamma (f) = \frac1F\,\sum_A \,
< e^{i 2\pi f\hat{\Omega}\cdot\Delta \vec{r}}\,
F_1^A (\hat{r}_1,\hat{\Omega},\psi) 
F_2^A (\hat{r}_2,\hat{\Omega},\psi) >_{\hat{\Omega},\psi}\,=\,
\frac{\Gamma (f)}{F}\,
\eeq
where $\Delta \vec{r} = \vec{r}_1 - \vec{r}_2$ is the separation 
vector between the two detector sites,  $F_i^A$ is the pattern 
function characterizing the response of the $i$-th detector ($i = 1,2$) 
to the $A = +,\times$ polarization, and the following notation
\beq 
< ... >_{\hat{\Omega},\psi}\,=\,
\int_{S^2}\,\frac{\ud \hat{\Omega}}{4 \pi}\,
\int_0^{2 \pi}\,\frac{\ud \psi}{2 \pi}\,( ... )
\eeq
has been introduced to indicate the average over the propagation 
direction $(\theta,\phi)$ and the polarization angle $\psi$. The 
normalization factor $F$ is given by:
\beq\label{normf}
F = \sum_A\,< F_1^A (\hat{r}_1,\hat{\Omega},\psi) 
F_2^A (\hat{r}_2,\hat{\Omega},\psi) >_{\hat{\Omega},\psi}
\,\mid_{1 \equiv 2}\,,
\eeq
where the notation $1 \equiv 2$ is a compact way to indicate that the 
detectors are coincident and coaligned and, if at least one of them is 
an interferometer, the angle between its arms is equal to $\pi$/2 
(L-shaped geometry). In this situation, by definition, $\gamma (f) = 1$. 
When the detectors are shifted apart (so there is a phase shift between 
the signals in the two detectors), or rotated out of coalignment (so the 
detectors have different sensitivity to the same polarization) it turns 
out that: $|\gamma (f)| < 1$.

The pattern functions (or orientation factors) of a GW detector can 
be written in the following form
\beq\label{patt}
F^A (\hat{r},\hat{\Omega},\psi) = \mbox{Tr}\,\{ D (\hat{r})\,
\varepsilon^A (\hat{\Omega},\psi) \}
\eeq
where the symmetric, trace-less tensor $D (\hat{r})$ describes the 
orientation and geometry of the detector located at $\vec{r}$. 

The tensor $D(\hat{r})$ depends upon the geometrical features of the 
detector. 
For instance, in the case of an 
 interferometer, indicating with $\hat{u}$ and $\hat{v}$ the 
unit vectors in the directions of its arms, one has:
\beq\label{dint}
D (\hat{r}) = \frac12\,\big\{\hat{u} (\hat{r}) \otimes \hat{u} (\hat{r}) 
\,-\,\hat{v} (\hat{r}) \otimes \hat{v} (\hat{r}) \big\}\,.
\eeq
In the case of  the lowest longitudinal mode of a cylindrical GW antenna with 
axis in the direction determined by the unit vector $\hat{l}$, one 
has
\beq\label{dcyl}
D (\hat{r}) = \hat{l} (\hat{r}) \otimes \hat{l} (\hat{r})\,-\,
\frac13 I\,,
\eeq
where $I$ is the unit matrix. 
Finally, in the case of the lowest five degenerate 
quadrupole modes ($m = -2, \ldots, +2$) of a spherical detector, the 
corresponding tensors are 
\beqn\label{dsph}
D^{\,(0)}\,(\hat{r}) &=& \frac{1}{2 \sqrt{3}}\,\big\{ 
e^+ (\hat{r}) \,+\,2\,g^+ (\hat{r}) \big\}
\,=\, \frac{1}{2 \sqrt{3}}\,\big\{ 2\,f^+ (\hat{r})\,-\,
e^+ (\hat{r}) \big\} \nonumber \\
D^{\,(+1)}\,(\hat{r}) &=& - \frac12\,g^\times (\hat{r}) \qquad 
\qquad D^{\,(-1)}\,(\hat{r}) = - \frac12\,f^\times (\hat{r}) \\
D^{\,(+2)}\,(\hat{r}) &=& \frac12\,e^+ (\hat{r}) \qquad \qquad 
\;\;\; D^{\,(-2)}\,(\hat{r}) = - \frac12\,e^\times (\hat{r}) \nonumber
\eeqn
where
\nbeqn
f^+ (\hat{r}) &=& \hat{m} (\hat{r}) \otimes \hat{m} (\hat{r})
\,-\,\hat{r} \otimes \hat{r} \qquad \qquad 
f^\times (\hat{r}) = \hat{m} (\hat{r}) \otimes \hat{r}\,+\,
\hat{r} \otimes \hat{m} (\hat{r}) \\
g^+ (\hat{r}) &=& \hat{n} (\hat{r}) \otimes \hat{n} (\hat{r})
\,-\,\hat{r} \otimes \hat{r} \qquad \qquad 
\;\;g^\times (\hat{r}) = \hat{n} (\hat{r}) \otimes \hat{r}\,+\,
\hat{r} \otimes \hat{n} (\hat{r})\,,
\neeqn
and $e^{+,\times} (\hat{r})$ are the tensors of Eq. (\ref{polten}) 
written in terms of the unit vectors $\hat{m} (\hat{r})$ and 
$\hat{n} (\hat{r})$ lying on the plane perpendicular to $\hat{r}$.
From these expressions for the tensors $D^{ij}$ and interpreting
each of the five modes of a sphere as a single detector, it is possible to 
show that in the case of coincident detectors one has:
\beq\label{corpat}
< F_1^A (\hat{r},\hat{\Omega},\psi) 
F_2^B (\hat{r},\hat{\Omega},\psi) >_{\hat{\Omega},\psi} 
\,=\,c_{12}\,\delta^{A B} \qquad \qquad \qquad (A, B = +, \times)
\eeq
where $c_{12}$ depends only on the geometry and the relative orientations 
of the two detectors. The corresponding values of $F$ (see Eq. (\ref{normf})) 
for the three different geometries considered (interferometer, cylindrical 
bar, sphere) are summarized in Table \ref{tab1}.
By introducing the following notation
\bdis
\Delta \vec{r} = d\,\hat{s} \qquad \qquad \delta = 2\,\pi\,f\,d\,,
\edis
where $\hat{s}$ is the unit vector along the direction connecting 
the two detectors and $d$ is the distance between them, it can be shown 
\cite{chr} that the overlap reduction function assumes the following 
form ($D_k = D (\hat{r}_k)$): 
\beq\label{gamten}
\gamma(f) = \rho_0 (\delta)\,D_1^{\,ij}\,D_{2\,ij}\,+\,\rho_1 (\delta)\,
D_1^{\,ij}\,D_{2\,i}^{\;k}\,s_j\,s_k\,+\,\rho_2(\eta)\,D_1^{\,ij}\,
D_2^{\,kl}\,s_i\,s_j\,s_k\,s_l
\eeq
where 
\beq\label{rhof}
\left[ 
\bary{c}
\rho_0 \\
\rho_1 \\
\rho_2
\eary
\right] (\delta) = \frac{1}{F \delta^2}\,
\left[
\bary{rrr}
 2 \delta^2  & -4 \delta   & 2 \\
-4 \delta^2  &  16 \delta  & -20 \\
    \delta^2 & -10 \delta  & 35
\eary
\right]\
\left[
\bary{c}
j_0 \\
j_1 \\
j_2
\eary
\right] (\delta)\,,
\eeq
with $j_k (\delta)$ the standard spherical Bessel functions:
\bdis
j_0 (\delta) = \frac{\sin{\delta}}{\delta}\,, \qquad 
j_1 (\delta) = \frac{j_0 (\delta) - \cos{\delta}}{\delta}\,, \qquad
j_2 (\delta) = 3\,\frac{j_1 (\delta)}{\delta}\,-\,j_0 (\delta)\,.
\edis
\begin{table}[!hpt] 
\bcen
\caption{\label{tab1}
The normalization factor $F$ for three different geometries of the 
detectors: interferometer (ITF), cylindrical bar (BAR), and sphere (SPH).
A $\star$ denotes entries that can be obtained from the symmetry of the 
table.} 

\vspace*{0.5cm}
$$
\begin{array}{|@{\qquad}c@{\qquad}||@{\qquad}c@{\qquad}|@{\qquad}c@{\qquad}
|@{~~} c c c @{~~} |}
\hline \rule{0ex}{3.5ex}
 & {\rm ITF} & {\rm BAR} & \multicolumn{3}{c|}{\rm SPH} \\ 
 & & & m\,=\,0 & m\,=\,\pm\,1 & m\,=\,\pm\,2 \\ 
\hline \hline \rule{0ex}{3ex} 
 {\rm ITF}  &  2/5 & \star & \star & \star & \star \\[5pt]
\hline \rule{0ex}{3ex}
 {\rm BAR}  &  2/5 & 8/15 & \star & \star & \star \\[5pt]
\hline \rule{0ex}{3ex}
 \qquad ~m\,=\,0 & 0 & 2 \sqrt{3}/15 & 2/5 & \star & \star \\[5pt]
 {\rm SPH} \quad m =\,\pm\,1 & 0 & 0 & 0 & 2/5 & \star \\[5pt]
 \qquad~~~~m\,=\,\pm\,2 & 2/5 & 2/5 & 0 & 0 & 2/5 \\[5pt]
\hline
\end{array}
$$
\ecen
\end{table}

\renewcommand{\theequation}{B.\arabic{equation}}
\setcounter{equation}{0}
\section{BBN bounds} 

In the case of minimal models the integrals determining 
the analytical expression of the BBN bound are given by:
\begin{eqnarray}
{\cal I}_{d} &=& z_{s}^{- 2 \beta}\,\left\{\,\frac{1}{54}\,
(z_s^2 + 6\,z_s + 18)\,-\,\frac{1}{108}\,\left(\,\frac{f_{\rm ns}}{f_{s}}\,
\right)^3\,\left[\,2\,(z_s^2 + 6\,z_s + 18) \right. \right. 
\nonumber \\
& & \qquad \qquad \qquad \left. \left.
-\,6\,z_s\,(z_s + 6)\,\ln{\frac{f_{\rm ns}}{f_s}}\,+
\,9\,z_s^2\,\ln^2{\frac{f_{ns}}{f_s}}\,\right]\,
\right\}\;,\nonumber\\[10pt]
{\cal I}_{s} &=& \frac{3}{2\,\beta\,(3 - 2 \beta)}\,+\,
\frac{z_s^{2 \beta - 6}}{2 \beta - 6}\,-\,\frac{z_s^{- 2 \beta}}
{2 \beta}\;.
\end{eqnarray}

In the case of non-minimal models the integrals determining 
the BBN bound are given by
\begin{eqnarray}
{\cal I}_{1} &=& A(\sigma, z_s)\,+\,B(\sigma, z_s)\,\ln{z_s}\,+\,
C(\sigma, z_s)\,\ln^2{z_s}\;,\nonumber\\[10pt]
{\cal I}_{2} &=& \frac{z_s^{-4}}{4}\,\left(\,z_s^{\sigma - 2}\,+\,
z_s^{2 + \sigma}\,\right)\,\left(\,z_s^{- 4}\,-\,z_r^{- 4}\,\right)\,
(1\,+\,\ln{z_s})^2\;,
\end{eqnarray}
where and $z_r\,=\,f_1/f_r$ and 
\begin{eqnarray}
A(\sigma, z_s) &=& - \frac{z_s^{2 \sigma}}{16\,(\sigma^2 - 4)^3}\,
\left\{\,13\,z_s^{- 2(2 + \sigma)}\,(\sigma^2 - 4)^3\,-\,4\,z_s^{-4}\,
(\sigma + 2)^3\,(2 \sigma^2 - 10 \sigma + 13) \right. \nonumber\\
& &\qquad \qquad \qquad \quad \left .
+\;4\,z_s^{- 4( 1 + \sigma)}\,(\sigma - 2)^3\,
(2 \sigma^2 + 10 \sigma + 13)\,-\,z_s^{- 2 \sigma}\,
(13 \sigma^6 - 172 \sigma^4 + 832 \sigma^2 - 1664)\,
\right\}\;, \nonumber \\[10pt] 
B(\sigma, z_s) &=& \frac{z_s^{2 \sigma - 4}}{4\,(\sigma^2 - 4)^2}\,
\left\{\,2\,(\sigma + 2)^2\,(2 \sigma - 5)\,-\,2\,z_s^{- 4 \sigma}\,
(\sigma - 2)^2\,(2 \sigma + 5)\,-\,5\,z_s^{2 \sigma}\,(\sigma^2 - 4)^2
\,\right\}\;, \\[10pt]
C(\sigma, z_s) &=& \frac{z_s^{4 - 2 \sigma}}{2\,(\sigma^2 - 4)}\,
\left\{\,2\,-\,z_s^{- 4 \sigma}\,(\sigma - 2)\,+\,\sigma\,
z_s^{- 2 \sigma}\,(\sigma^2 - 4))\right\}\;. \nonumber
\end{eqnarray}

\end{appendix}

\end{document}